\documentclass[12pt,preprint]{aastex}
\usepackage{natbib}

\begin{document}

\title{Gravitational Instabilities in Two-Component Galaxy Disks with
Gas Dissipation}

\author{Bruce G. Elmegreen}
\affil{IBM T. J. Watson Research Center, 1101 Kitchawan Road, Yorktown
Heights, New York 10598 USA} \email{bge@us.ibm.com}

\begin{abstract}
Growth rates for gravitational instabilities in a thick disk of gas and
stars are determined for a turbulent gas that dissipates on the local
crossing time. The scale heights are derived from vertical equilibrium.
The accuracy of the usual thickness correction, $(1+kH)^{-1}$, is
better than 6\% in the growth rate when compared to exact integrations
for the gravitational acceleration in the disk. Gas dissipation extends
the instability to small scales, removing the minimum Jeans length.
This makes infinitesimally thin disks unstable for all Toomre-$Q$
values, and reasonably thick disks stable at high $Q$ primarily because
of thickness effects. The conventional gas+star threshold, $Q_{\rm
tot}$, increases from $\sim1$ without dissipation to 2 or 3 when
dissipation has a rate equal to the crossing rate over a perturbation
scale. Observations of $Q_{\rm tot}\sim2-3$ and the presence of
supersonic turbulence suggest that disks are unstable over a wide range
of scales. Such instabilities drive spiral structure if there is shear
and clumpy structure if shear is weak; they may dominate the generation
of turbulence. Feedback regulation of $Q_{\rm tot}$ is complex because
the stellar component does not cool; the range of spiral strengths from
multiple arm to flocculent galaxies suggests that feedback is weak.
Gravitational instabilities may have a connection to star formation
even when the star formation rate scales directly with the molecular
mass because the instabilities return dispersed gas to molecular clouds
and complete the cycle of cloud formation and destruction. The mass
flow to dense clouds by instabilities can be 10 times larger than the
star formation rate.
\end{abstract}

\section{Introduction}

The stability of a galactic disk with gas and stars was first
considered in the context of stellar waves by \cite{lin66} and in the
context of cloud formation by \cite{jog84}. The importance of cold gas
in the general stability of a combined disk was recognized by
\cite{toomre64}, who also considered the effects of thickness and
particle dynamics for the stars, as did \cite{kal65}. Two-component
disks of finite thickness were considered in more detail for the spiral
wave problem by \cite{kato72} and \cite{bertin88} and in the general
case by \cite{romeo92}. \cite{rafikov01} investigated multiple
components.

With these two-component analyses came stability parameters that
determined in absolute terms when a disk was stable. For one stellar
component, this is the \cite{toomre64} value for the minimum velocity
dispersion in the radial direction, written now as a dimensionless
parameter $Q$. \cite{goldreich65} evaluated a maximum density for the
stability of a one-component disk. Following the two-fluid analysis by
\cite{jog84}, \cite{e95} and \cite{jog96} determined a combined $Q$ for
a two-fluid disk. A simpler form for the combined $Q$ was written by
\cite{wang94}, and recently improved by \cite{romeo11}.
\cite{rafikov01} discussed disk stability using the dispersion relation
at zero frequency, which is a function of wavenumber, and gave an
expression that was compared to observations for spiral galaxies and
dwarfs by \cite{leroy08}. Leroy et al. evaluated the Rafikov expression
assuming a range of likely wavenumbers, and showed that most radii in
all of the galaxies they studied were marginally stable. This common
feature suggests that galaxies regulate their stability by varying the
velocity dispersions of gas and stars through star formation feedback
and stellar wave scattering.

An important component of disk stability that has been overlooked in
these studies is the dissipation of energy in the interstellar medium.
Usually the gas is assumed to be isothermal or polytropic, which has a
certain level of dissipation by assumption, but the nature of this
dissipation is unrealistic as it implies an ever-present source of
energy to accompany any decompression. On average, lower-density gas is
warmer than higher-density gas, but this relationship is established in
the real ISM by specific energy sources and sinks that do not
automatically heat up the gas if it is driven to a lower density by
gravity. A time-dependent energy equation should be used instead of an
adiabatic pressure-density relation. In equilibrium, this energy
equation states that heating balances cooling, but out of equilibrium,
when a spiral arm or cloud begins to form in an instability, there is
usually a net cooling from dissipation in turbulence. Only after the
spiral or cloud make stars will the energy loss be returned, but this
is long after the instability has grown. Recent computer simulations
suggest that the primary role of star formation feedback in a galaxy
model is to break apart the dense clouds that form; turbulence on
larger scales is driven by gravitational instabilities (Bournaud et al.
2010, Hopkins et al. 2011, see also Elmegreen et al. 2003). This means
that turbulent dissipation accompanies the growth of gravitational
instabilities.

The purpose of this paper is to introduce the energy equation to the
interstellar component of the stability problem for a gas and star
disk.  We also consider thickness effects in some detail, checking the
approximation often used for stability analyses with an exact
integration for the gravitational acceleration in a three dimensional
disk. Then we discuss corrections to the commonly used stability
thresholds that arise from gaseous energy dissipation.  The main point
is that energy dissipation changes the dispersion relation for disk
instabilities in a fundamental way, removing the Jeans length from the
problem in some cases, and giving residual instabilities at high
wavenumbers even when the Toomre Q parameters are large. Disk thickness
is a more effective stabilizer than interstellar turbulent pressure.

The results have implications for the interpretation of observations,
primarily in suggesting a better stability threshold when dissipation
is important, and they have implications for simulations of galaxy
disks. In simulations, an equation of state for the gas has to be
considered to connect density changes with pressure changes.  If this
equation of state is too simple, particularly if it does not reproduce
the expected rapid dissipation of turbulent energy in perturbations,
then the simulation is probably more stable than a real galaxy -- at
least on small scales.

Another application is to regions of galaxies where the usual stability
parameters suggest a high degree of stability, such as the far-outer
regions of disks, or in dwarf galaxies
\citep{hunter96,meurer96,vanzee97,hunter98}. The present results
suggest that these regions can still be unstable at a low level as
perturbations in density lead to a simultaneous increase in
self-gravity and a decrease in gaseous velocity dispersion because of
dissipation.  Such a region will collapse if dissipation operates on
about a crossing time, because the pressure does not increase fast
enough to overcome gravity.

In what follows, the star+gas dispersion relation is derived in Section
\ref{sect:eq}, a simple example including turbulent dissipation but
without stars is discussed in Section \ref{jeans}, and the star+gas
relation is evaluated in Section \ref{sect:sol}. Gas and star growth
rates are considered separately in Section \ref{gss} and together in
Section \ref{gst}. The ratio of perturbed stellar and gaseous column
densities is in Section \ref{ratio}. Section \ref{thresh} discusses
dissipation corrections to the popular gas+star $Q$ thresholds that are
in the literature. Section \ref{sect:azim} examines thresholds for
azimuthal instabilities. A discussion of several implications of this
work is in Section \ref{disc}, including the greater role played by
instabilities when dissipation is included (Sect. \ref{sect:perv}),
feedback regulation of disk stability (Sect. \ref{sect:feed}), and star
formation (Sect. \ref{sf}). A summary is in Section \ref{sect:sum}.

The scale height of the disk enters the thickness correction factor in
the gravitational forcing term. Scale heights are evaluated in a
self-consistent way in Appendix A and used for the solutions of the
dispersion relation in Section \ref{sect:sol}. The accuracy of the
thickness correction factor is determined in Appendix B.

\section{Dispersion Relation}\label{sect:eq}

The equations for perturbed radial and azimuthal motions of gas in an
axisymmetric galaxy disk are \citep[e.g.,][]{rafikov01}
\begin{equation}
{{\partial v_{\rm g,r}}\over{\partial t}}-2\Omega v_{\rm g,\theta} =
-{{\partial \phi} \over{\partial r}}-{1\over \rho_{\rm g,0}}{{\partial
P}\over {\partial r}},\label{eq1}
\end{equation}
\begin{equation}
{{\partial v_{\rm g,\theta}}\over{\partial t}}-2B v_{\rm g,r}
=0.\label{eq2}
\end{equation}
Here, $v_{\rm g,r}$ and $v_{\rm g,\theta}$ are the perturbed radial and
azimuthal gas velocities, $\phi$ is the perturbed gravitational
potential, $P$ is the perturbed pressure and $\rho_{\rm g,0}$ is the
unperturbed gas density. The angular rotation rate is $\Omega$, and $B$
is the Oort constant of rotation.

Perturbations of the form $e^{i(kr-\omega_{\rm i} t)}$ are typically
introduced. The oscillation frequency is denoted by $\omega_{\rm i}$ to
distinguish it from a growth rate $\omega$ introduced later, i.e.,
$\omega=-i\omega_{\rm i}.$ The equations of motion are combined with
the definition of the epicyclic frequency, $\kappa^2=-4\Omega B$, to
give
\begin{equation}
v_{\rm g,r}=-{{k\omega_{\rm i}}\over{\kappa^2-\omega_{\rm
i}^2}}\left(\phi+P/\rho_{\rm g,0}\right).\label{eq:vgr}\end{equation}

The stars were originally treated by \cite{toomre64}, \cite{kal65}, and
\cite{lin66} using the collisionless Boltzmann equation for the
phase-space distribution function. After analogous reduction, the
stellar equation becomes \citep[see also][]{rafikov01}
\begin{equation}
v_{\rm s,r}=-{{k\omega_{\rm i}\phi}\over{\kappa^2-\omega_{\rm i}^2}}
{\cal F}(\omega_{\rm i}/\kappa,[k\sigma_{\rm s}/\kappa]^2)
\label{eq:vsr}\end{equation} where
\begin{equation}
{\cal F}(s_{\rm i},q^2)={{1-s_{\rm i}^2}\over{\sin (\pi s_{\rm
i})}}\int_0^\pi \exp\left( -q^2 [1+\cos\xi]\right)\sin(s_{\rm
i}\xi)\sin\xi d \xi\label{eq:f}
\end{equation} \citep{binney08}. The notation here is that
$s_{\rm i}$ is the dimensionless oscillation frequency, $\omega_{\rm
i}/\kappa$, to distinguish it from a dimensionless growth rate,
$s=-is_{\rm i}$, introduced below.  The radial velocity dispersion of
stars in a presumed Schwarzschild distribution function is $\sigma_{\rm
s}$; we define dimensionless wavenumber $q=k\sigma_{\rm s}/\kappa$.

The left-hand side of Figure 1 shows ${\cal F}(s_{\rm i},q^2)$ as a
function of dimensionless squared wavenumber $q^2$ for various
dimensionless oscillation frequencies $s_{\rm i}$. When $s_{\rm i}$ is
small,
\begin{equation}
{\cal F}(s_{\rm i},q^2)\approx {{1-s_{\rm i}^2}\over{1-s_{\rm
i}^2+q^2}}.\label{eq:approx}
\end{equation} In the figure, the solid lines of various colors are the
exact values of ${\cal F}$ and the dashed lines with the corresponding
colors are the approximate values. \cite{toomre64} recognized this
similarity for $s_{\rm i}=0$.  The deviation between the two functions
is too small to be seen in most of the figure. A blow-up insert shows
how similar the solutions are for $s_{\rm i}=0.1$ (green curves),
$s_{\rm i}=0$ (blue curves), and $s_{\rm i}=0.5$ (magenta curves). The
similarity between the exact and approximate curves for small $s_{\rm
i}$ means that the instability condition for a collisionless stellar
fluid, which is derived from the dispersion relation in the limit when
$s_{\rm i}=0$, is nearly the same as the instability condition for a
fluid having the same velocity dispersion and mass column density. This
is why the condition $Q_s=\kappa\sigma_{\rm s}/\left(3.36 G\Sigma_{\rm
s,0}\right)<1$ for stars has the constant 3.36 in it, while the
condition $Q_g=\kappa\sigma_{\rm g}/\left(\pi G\Sigma_{\rm
g,0}\right)<1$ for gas has the constant $\pi$.  This constant, 3.35828,
is the maximum value of $2\pi{\cal F}(0,q^2)/q$ over variations in $q$.

Equations (\ref{eq:vgr}) and (\ref{eq:vsr}) contain the radial
velocity, gravitational potential, and pressure. The latter two are
related to the mass column density.  This column density is related to
velocity through the continuity equation, which is, for perturbed
variables,
\begin{equation}
\omega_{\rm i}\Sigma_{\rm g}=kv_{\rm g,r}\Sigma_{\rm g,0}.
\label{eq:sigmag}\end{equation}
\begin{equation}
\omega_{\rm i}\Sigma_{\rm s}=kv_{\rm s,r}\Sigma_{\rm s,0}.
\label{eq:sigmas}\end{equation}

Column density is related to pressure through a time-dependent energy
equation. The equation for perturbed energy, when multiplied by
$\gamma-1$ for adiabatic index $\gamma$, becomes an equation for
perturbed pressure changes:
\begin{equation}
{{\partial P}\over{\partial t}} = \left({{\gamma P_0}\over{\rho_{\rm
g,0}}}\right)\left({{\partial \rho} \over{\partial t}} \right) +
(\gamma-1)\left(\Gamma-\Lambda\right).\label{eq:p}
\end{equation} The perturbed rate of increase of energy per unit volume is
$\Gamma$ and the perturbed rate of decrease from radiative cooling and
other effects is $\Lambda$. We consider that star formation is mostly
triggered by the disk instabilities that we want to study, so that
before star formation begins, there is very little perturbation in the
heating rate. Perturbed heating might come from background supernovae,
for example, and depend on the gas density so that the rate changes
when the density changes. However, the heating rate from supernovae
depends very weakly on density \cite[as $\rho^{-0.1}$, as may be
derived from][]{cioffi}, so we set the perturbed $\Gamma$ equal to
zero. This means that the background turbulent heating rate per unit
volume does not increase much as the density increases during the
growth of a perturbation.

Cooling comes mostly from shocks in an interstellar medium that is
supersonically turbulent. The cooling rate should be larger where the
kinetic energy density is larger, which is where the pressure is also
larger.  To the extent that we can approximate the momentum flux for a
supersonically turbulent fluid with the pressure term, $P$, we can also
approximate the dissipation time for the corresponding energy density
as some number, $\delta^{-1}$,  of crossing times in the perturbed
region.  This crossing time dependence follows from numerical
simulations in \cite{maclow98} and \cite{stone98}. Because the crossing
rate is $\sigma_{\rm g}k$, we write the dissipation rate as
\begin{equation}
(\gamma-1)(-\Lambda) = -\delta \sigma_{\rm g}kP.
\label{eqheat}\end{equation} It is significant that the perturbed
cooling rate peaks at the same position as the perturbed pressure,
rather than at the position of peak pressure gradient.  Then the time
dependence of the pressure in equation (\ref{eq:p}) is in phase with
the pressure itself,
\begin{equation}
{{\partial P}\over{\partial t}} = \left({{\gamma P_0}\over{\rho_{\rm
g,0}}}\right)\left({{\partial \rho} \over{\partial t}} \right) -\delta
\sigma_{\rm g}kP.\label{eq:p2}
\end{equation}
This phase coherence means that the pressure has a component to the
time dependence that is exponential. There are no purely oscillatory
solutions as there are for polytropic or isothermal gasses. Thus we
need solutions of the form $e^{ikr+\omega t}$ for real $\omega$. We
also expect that for self-gravitating media there will be growing
(unstable) solutions, in which case $\omega>0$. Note that the
derivation of ${\cal F}$ also requires the real part of the growth term
to be positive so that an integral converges at negative infinite time
\citep[][Appendix 6A]{binney08}. We identify $\omega$ as the growth
rate of the gas+star gravitational instability.

The reduction factor ${\cal F}$ includes solutions where the time
dependence is exponential. Substituting $is$ for $s_{\rm i}$ in
equation (\ref{eq:f}), we obtain for growing perturbations,
\begin{equation}
{\cal F}_{\rm growth}(s,q^2)={{1+s^2}\over{{\rm sinh}( \pi
s)}}\int_0^\pi \exp\left( -q^2 [1+\cos\xi]\right) {\rm sinh}
(s\xi)\sin\xi d \xi\label{eq:f2}
\end{equation}
where $s$ is the dimensionless growth rate, $s=\omega/\kappa$. The
right-hand side of Figure 1 shows ${\cal F}$ for growing solutions. An
excellent approximation is equation (\ref{eq:approx}), which is now
written
\begin{equation}
{\cal F}_{\rm growth}(s,q^2)\approx {{1+s^2}\over{1+
s^2+q^2}}.\label{eq:approx2}
\end{equation}
This approximation, shown by the dashed lines, is closer to the exact
growth solution than equation (\ref{eq:approx}) was to the oscillatory
solution, equation (\ref{eq:f}).

The dissipation term in equation (\ref{eq:p}) has another important
effect because it removes the minimum unstable length, or Jeans'
length, from the problem for an infinitesimally thin disk. When
$\delta\ne0$, arbitrarily small perturbation lengths can dissipate
turbulent energy and drive an instability. It also removes a stability
condition: gas perturbations that always cool are always unstable. Then
there is no minimum gas column density for instability, nor is there
anything analogous to a Toomre $Q$ condition for an infinitesimally
thin disk. Note that zero-order heating and cooling are assumed to be
in equilibrium. The turbulent dissipation of zero-order pressure is
balanced by energy input from background sources, such as supernovae.
But to the extent that perturbed pressures driven by self-gravity
always dissipate internal turbulent energy faster than they gain
turbulent energy from background sources, there is an instability at
all length scales. In practice, that means all length scales where the
gas is supersonically turbulent.  We return to this point in Section
\ref{sect:sol} where the effects of disk thickness lead to a stability
condition again.

At this point we introduce perturbation variables that clearly indicate
their spatial phase dependence.  We let perturbed pressure, density,
column density and gravitational potential vary as $\cos kr$, and we
let both velocities vary as $\sin kr$. All perturbed quantities varying
in time as $e^{\omega t}$. It follows that equations (\ref{eq:vgr}) and
(\ref{eq:vsr}) become
\begin{equation}
v_{\rm g,r}={{k\omega}\over{\kappa^2+\omega^2}}\left(\phi+P/\rho_{\rm
g,0}\right), \label{eq:vgr2}\end{equation}
\begin{equation}
v_{\rm s,r}={{k\omega\phi}\over{\kappa^2+\omega^2}} {\cal F}_{\rm
growth}(\omega/\kappa,[k\sigma_{\rm s}/\kappa]^2).
\label{eq:vsr2}\end{equation} The continuity equations are
\begin{equation}
\omega\Sigma_{\rm g}=-kv_{\rm g,r}\Sigma_{\rm g,0},
\label{eq:sigmag2}\end{equation}
\begin{equation}
\omega\Sigma_{\rm s}=-kv_{\rm s,r}\Sigma_{\rm s,0}.
\label{eq:sigmas2}\end{equation} The energy (pressure) equation
simplifies to
\begin{equation}
P={{\gamma P_{\rm 0}\Sigma_{\rm g}}\over{\Sigma_{\rm g,0}}}
\left({\omega\over {\omega+\delta \sigma_{\rm g}k}}
\right).\label{eq:P2}\end{equation} Pressure goes to zero as $\omega$
goes to zero if $\delta\ne0$. This is why pressure does not enter the
stability condition and why there is no minimum length in the case of
turbulent dissipation like that given by equation (\ref{eqheat}).

The final link is between column density and gravitational potential.
This comes from Poisson's equation for the gravitational potential and
has no time dependence.  The equation is three-dimensional, so the
perturbed potential depends on the distribution of perturbed density
perpendicular to the plane. This leads to correction factors for the
mass column densities if only the two-dimensional forcing is
considered. \cite{toomre64} used a correction factor of
$(1-e^{-kH})/(kH)$ for a one-component uniform disk between $z=\pm H$,
while \cite{vandervoort70} used $(1+kH)^{-1}$ from a more detailed
analysis. These two factors are similar, as is the ${\cal J}$ factor in
Shu (1968, see Vandervoort 1970).  The use of analogous correction
factors in the two-component case was discussed by \cite{romeo92}. He
concluded that the correction factors cannot generally be applied
independently to each component, but if the vertical distributions are
somewhat uncoupled, then that approximation should be all right. In
this case the result is a perturbed gravitational potential that
depends on the perturbed gaseous and stellar column densities with
separate and independent correction factors,
\begin{equation}
\phi={{-2\pi G}\over k}\left({{\Sigma_{\rm g}}\over{1+kH_{\rm g}}} +
{{\Sigma_{\rm s}}\over{1+kH_{\rm s}}}\right).\label{eq:grav}
\end{equation}
We use this approximation here.  In Appendix \ref{sect:height}, the two
scale heights $H$ are evaluated from the equations of vertical
equilibrium. In Appendix \ref{correct}, we check the thickness
approximation in equation (\ref{eq:grav}) by direct integration over
the combined density distribution to determine the gravitational
acceleration from a perturbation. The result suggests that the
$(1+kH)^{-1}$ correction factors for column density are too low by
$\sim12$\% at $kH\sim1-10$, but this is the largest error introduced by
the thickness approximation for a wide range of $\Sigma_{\rm
g,0}/\Sigma_{\rm s,0}$ and other parameters.

With equations (\ref{eq:vsr2}), (\ref{eq:sigmas2}), and (\ref{eq:grav})
we can eliminate $\phi$ and $v_{\rm s,r}$ to get an equation for
perturbed $\Sigma_{\rm s}/\Sigma_{\rm g}$:
\begin{equation}
{{\Sigma_{\rm s}\left(1+kH_{\rm g}\right)}\over {\Sigma_{\rm
g}\left(1+kH_{\rm s}\right)}} ={{Y_{\rm s}}\over {1-Y_{\rm
s}}},\label{eq:ss}\end{equation} where
\begin{equation}
Y_{\rm s}={{2\pi Gk\Sigma_{\rm s,0} {\cal F}_{\rm
growth}(s,q^2)}\over{\left(\omega^2+\kappa^2\right) \left(1+kH_{\rm
s}\right)}}\end{equation} and $s=\omega/\kappa$, $q=k\sigma_{\rm
s}/\kappa$, as defined above.

Now equations (\ref{eq:vgr2}), (\ref{eq:sigmag2}), (\ref{eq:P2}),
(\ref{eq:grav}), and (\ref{eq:ss}) combine to give the dispersion
relation,
\begin{equation}
1=Y_{\rm g}+Y_{\rm s}\label{dr:y}\end{equation} where
\begin{equation}
Y_{\rm g}={{2\pi Gk\Sigma_{\rm
g,0}}\over{\left(\omega^2+\kappa^2+\sigma_{\rm
g}^2k^2\omega/\left[\omega+\delta \sigma_{\rm
g}k\right]\right)\left(1+kH_{\rm g}\right)}},\end{equation} and the
square of the unperturbed gas velocity dispersion is $\sigma_{\rm
g}^2=\gamma P_0/\rho_{\rm g,0}$. With dimensionless variables, this
dispersion relation is
\begin{equation}
1={{2qS}\over {Q_{\rm g}\left(1+s^2+q^2S^2s/[s+\delta qS]\right)
\left(1+q{\hat H}_{\rm g}\right)}} + {{2q{\cal F}_{\rm
growth}(s,q^2)}\over {Q_{\rm s}\left(1+s^2\right)\left(1+q{\hat H}_{\rm
s}\right)}},\label{dr2}\end{equation} where $s=\omega/\kappa$,
$q=k\sigma_{\rm s}/\kappa$, $S=\sigma_{\rm g}/\sigma_{\rm s}$, ${\hat
H}_{\rm g}=H_{\rm g}\kappa/\sigma_{\rm s}$, ${\hat H}_{\rm s}=H_{\rm
s}\kappa/\sigma_{\rm s}$, $Q_{\rm g}=\kappa\sigma_{\rm g}/(\pi
G\Sigma_{\rm g,0})$ and $Q_{\rm s}=\kappa\sigma_{\rm s}/(\pi
G\Sigma_{\rm s,0})$. This expression is analogous to that in
\cite{jog84}, \cite{romeo92} and \cite{rafikov01}. It differs in our
consideration of turbulent gas dissipation and in our use of growth
solutions for the reduction factor ${\cal F}$.

In \cite{rafikov01}, the instability condition for a gas+star fluid was
that the right-hand side of equation (\ref{dr2}), written with
$\delta=0$ and $H_{\rm g}=H_{\rm s}=0$ in his case, had to exceed 1 in
the limit of $s=0$ for some value of wavenumber $q$. Stability requires
the right hand side to be less than 1 at $s=0$ for all wavenumbers.  If
$\delta=0$, we have these conditions too, because then the gas term
varies as $qS/(1+q^2S^2)$ which has a maximum value of 0.5 at $qS=1$,
and then sufficiently small $Q_{\rm g}$ makes the right hand side
larger than 1. When $\delta\ne0$, however, the gas term varies directly
with $q$ when $s=0$ and there is no maximum value. The right hand side
can exceed 1 for $s=0$ even for very large $Q_{\rm g}$, which is
normally stable, if $q$ is large enough. For an infinitesimally thin
disk, we need $q>Q_{\rm g}/2S$ or $k>\kappa^2/(2\pi G\Sigma_{\rm g,0})$
for instability in the pure gas case ($\Sigma_{\rm {s,0}}=0$) if there
is perturbed gaseous energy dissipation of any magnitude, i.e.,
$\delta>0$. This is the same as the instability condition for a
infinitesimally thin pressure-less fluid, as derived by Toomre (1964,
eqn. 21).

\section{Jeans Relation with Dissipation}\label{jeans}

The role of turbulent dissipation in the standard Jeans analysis can be
seen from simpler equations involving only the gas. Equation
(\ref{dr:y}) with the first term on the right is for gas only, but that
also contains the effects of rotation and finite thickness. Solutions
to this equation will be shown in the next section. An even simpler
form is to write it for an infinitesimally thin disk without rotation
in the form
\begin{equation}
\omega^2=2\pi G \Sigma_{\rm
g}k-{{\sigma_g^2k^2\omega}\over{\omega+\delta\sigma_{\rm g}k}}.
\end{equation}
For the present discussion, we make this equation dimensionless by
defining the unit of frequency-squared to be $2\pi G\Sigma k_{\rm J}$
and the unit of wavenumber-squared to be $k_{\rm J}^2= 2\pi
G\Sigma_{\rm g}/\sigma_{\rm g}^2$; $k_{\rm J}$ is the conventional
Jeans wavenumber. The two-dimensional Jeans equation with this
normalization is
\begin{equation}
\omega^2=k-{{k^2\omega}\over{\omega+\delta k}}. \end{equation} This is
a cubic equation,
\begin{equation}
\omega^3+\omega^2k\delta+\omega(k^2-k)-k^2\delta=0,
\end{equation}
that can be solved analytically, although the steps are complicated
with terms in $\delta$ up to order 6.

Figure \ref{jeans_w_delta} shows numerical evaluations of this
analytical solution as solid lines for $\delta=0$, 0.5 and 1.  The
usual dispersion relation with a wavenumber limit for growth at
$k/k_{\rm J}=1$ is reproduced in the case of no dissipation,
$\delta=0$.  With dissipation, the instability persists at high
wavenumber. The origin of this instability is uncompensated gas cooling
on $\delta^{-1}$ crossing times, aided by self-gravitational
contraction. In other words, turbulent shocks dissipate energy and
self-gravity then leads to collapse.

Figure \ref{jeans_w_delta} also shows as dashed lines solutions with a
dissipation rate proportional to $\delta \sigma_{\rm g,0}k^{1/2}$,
which is appropriate for a turbulent gas with a velocity dispersion
proportional to the square root of the size: $\sigma_{\rm g}\propto
k^{-1/2}$. There is still instability at high $k$, but the dissipation
rate for each $k$ is lower when there are slower motions at that $k$,
and so the growth rate is lower too.  The rest of this paper considers
only solutions with $\sigma_{\rm g}$ independent of $k$, but the same
shift toward lower growth rate at high $k$ would be present in the full
solutions too if $\sigma_{\rm g}$ increased with size.

\section{Evaluation of the Dispersion Relation}\label{sect:sol}

\subsection{Gas and Stars Considered Separately}\label{gss}

The first term on the right of equation (\ref{dr2}), when set equal to
1, is the dispersion relation for pure gas, and the second term is the
dispersion relation for pure stars. Combined, they are the dispersion
relation for stars+gas.  We show each of these in the next 2 figures.
The scale heights come from Appendix A.

Figure \ref{rafikov_dispersion_gasstars} shows the pure-gas dispersion
relation on the left and the pure-star dispersion relation on the
right. Both peak at dimensionless wavenumber $q\sim1-2$ and both have
growth rates that increase with decreasing $Q$. In the case of pure
gas, the effect of the dissipation term $\delta q S$ in equation
(\ref{dr2}) is to stretch the range of instabilities to larger
wavenumber. When $\delta=0$, the growth rate goes to zero at $q\sim2$
for $Q_g=0.5$ and at $q=3.7$ for $Q_g=0.3$.

In the case of pure stars (Figure \ref{rafikov_dispersion_gasstars}
right), the full growth solutions are plotted as solid blue curves.
These include the scale height term, $1+q{\hat H}_{\rm s}$ with a value
of ${\hat H}_{\rm s}$ that increases from 0.34 at $Q_{\rm s}=0.35$ for
the top blue curve, to 0.39, 0.49, and 0.58 at $Q_{\rm s}=0.4,$ 0.5,
and 0.6 for blue curves with decreasing heights. They also include the
${\cal F}_{\rm growth}$ term. For comparison, the dashed green curve
has $Q_{\rm s}=0.5$ and no scale height term. The growth is then faster
because the full stellar column density is used, not the value reduced
by $1+q{\hat H}_{\rm s}$.  For the value of ${\hat H}_{\rm s}=0.49$ in
this $Q_{\rm s}=0.5$ case,  $1+q{\hat H}_{\rm s}\sim1.49$ at $q=1$ near
the peak, and this makes the growth rate like the one for a full
dispersion relation with $Q_{\rm s}=0.5/1.49=0.34$ at the same $q$.
Figure \ref{rafikov_dispersion_gasstars} includes this case as the top
blue curve ($Q_{\rm s}=0.35$), for confirmation; the rest of that blue
curve differs from the $Q_{\rm s}=0.5$, ${\hat H}_{\rm s}=0$ case
because the scale height term also has a $q$ dependence.  That is, at
increasing $q$, the term $1+q{\hat H}_{\rm s}$ becomes larger and the
gravitating effective column density, $\Sigma_{\rm s}/(1+k{\hat H}_{\rm
s})$, smaller, making the disk less unstable than the $H_{\rm s}=0$
solutions.

In a second comparison, Figure \ref{rafikov_dispersion_gasstars}
(right, red dotted line) also shows the dispersion relation for $Q_{\rm
s}=0.5$ and ${\hat H}_{\rm s}=0$ in the case where the oscillatory
solution for the equations of motion are used, with ${\cal F}$ from
equation (\ref{eq:f}). This solution has no physical meaning because at
such low $Q_{\rm s}$ the solutions are unstable and not wave-like, but
it does show the difference that comes from using oscillatory solutions
for the reduction factor, rather than growth solutions.  The
oscillatory solutions vary approximately as ${\cal F}\sim (1-s_{\rm
i}^2)/(1-s_{\rm i}^2+q^2)$, which is smaller than the growth solutions
which vary as ${\cal F}_{\rm growth}\sim(1+s^2)/(1+s^2+q^2)$.  This is
why the red dotted line in Figure \ref{rafikov_dispersion_gasstars} is
lower than the green dashed line for $q>1$.

Figure \ref{rafikov_dispersion_gas2} shows another case with pure gas,
this time for $Q_{\rm g}>1$, which is normally stable. Here the
destabilizing effect of the dissipation term, $\delta qS$, can be seen.
The solid blue curves are a sequence of increasing $Q_{\rm g}=1$, 1.2,
and 1.4, from left to right, with dissipation $\delta=1$. The dashed
blue curve is the same as the $Q_{\rm g}=1$ curve but with smaller
dissipation, $\delta=0.5$, and so smaller growth rate. The solid red
curve has the same $Q_{\rm g}=1$ and $\delta=1$ as the top blue curve,
but it has a larger scale height obtained by setting $S=0.4$ instead of
0.2 in the other curves.

Figure \ref{rafikov_dispersion_gas2} indicates that although the
dissipative disk is unstable for some $Q_{\rm g}>1$, the growth
decreases quickly as $Q_{\rm g}$ increases.  Eventually the disk
becomes stable at large enough $Q_{\rm g}$ if the scale height term is
included. Recall in Section \ref{sect:eq} we said that infinitesimally
thin dissipative disks are unstable for any $Q_{\rm g}$ as long as
$q>Q_{\rm g}/2S$. With finite thickness, there can be stability at high
$Q_{\rm g}$. Setting $s=0$ in equation (\ref{dr2}) and using only the
first term on the right, which is for gas only, we see that if
$2qS/\left(Q_{\rm g}\left[1+q{\hat H}_{\rm g}\right]\right)>1,$ then
that equation admits unstable solutions with $s^2>0$. This condition
rearranges to
\begin{equation}
q>{{Q_{\rm g}}\over{2S-Q_{\rm g}{\hat H}_{\rm g}}}.
\label{qgt}\end{equation} With ${\hat H}_{\rm g}=0$ we get $q> Q_{\rm
g}/2S$ again, but with ${\hat H}_{\rm g}>0$, we also need $Q_{\rm
g}<2S/{\hat H}_{\rm g}$. Unwrapping all the dimensionless variables,
this condition is $Q_{\rm g}<2^{0.5}$ if $H=\sigma_{\rm g}^2/(\pi
G\Sigma_{\rm g,0})$ for a one-component dissipative disk.

\subsection{Gas and Stars Together}\label{gst}

Solutions to the combined gas+star dispersion relation (\ref{dr2}) are
shown in Figure \ref{rafikov_dispersion_sg}. The standard one-component
solutions without thickness corrections ($H_{\rm i}=0$) are shown for
comparison in the left-hand panel as the multicolored dotted line at
small $q$ and $s$. This line is made from the superposition in
different colors of a pure-star solution with $Q_{\rm s}=0.9$ and
$Q_{\rm g}=\infty$, a pure-gas solution with $Q_{\rm s}=\infty$,
$Q_{\rm g}=0.9$, and $\delta=0$, and a gas+star solution with gas and
stars equivalent, using $Q_{\rm s}=1.8$, $Q_{\rm g}=1.8$, $\delta=0$,
and $S=1$.  They are expected to have the same dispersion relation and
they do.  Also shown in the left-hand panel as red curves are two
pure-gas solutions ($Q_{\rm s}=\infty$, $Q_{\rm g}=0.9$) with different
dissipation rates, $\delta=0.1$ (lower curve) and $\delta=0.2$.  The
extension to high $q$ because of dissipation was also shown in Figures
\ref{jeans_w_delta}-\ref{rafikov_dispersion_gas2}. The solid blue
curves are two gas+star one-component-like cases with $Q_{\rm s}=0.9$,
$Q_{\rm g}=0.9$, $\delta=0$ and $S=1$, where in the top curve there are
no thickness corrections (${\hat H}_{\rm i}=0$) and in the bottom curve
there are. Thickness corrections lower the growth rate by about a
factor of 60\% in this case, where ${\hat H}_{\rm s}={\hat H}_{\rm
g}=0.44$, because $(1+q{\hat H}_{\rm i})^{-1}$=0.6 at $q=1.5$ near the
peaks of these curves.  Finally, another comparison with two different
dissipation rates is shown by the dashed blue lines but now with
gas+stars together. Here $Q_{\rm s}=1$, $Q_{\rm g}=0.5$, and $S=0.5$ in
both cases, while $\delta=0$ for the bottom curve and $\delta=0.5$ for
the top curve. The extension to high $q$ because of dissipation is
present.

A range of values for $Q_{\rm s}$ and $Q_{\rm g}$ are shown for the
two-component dispersion relation with thickness corrections in the
middle panel. The dissipation parameter is fixed at $\delta=0.5$ and
the velocity dispersion ratio is $S=0.5$. These four parameters all
enter into the dispersion equation (\ref{dr2}) directly. Also fixed is
the ratio of the rotation speed to the stellar dispersion ${\hat
v}_{\rm rot}={\hat v}_{\max}=8$, the rotation curve slope $\alpha=0.4$,
and the dark matter to disk ratio, $D=1$, which enter into the scale
height equations (\ref{drdz2}) and (\ref{drdzs2}).  The sequence of
blue curves with decreasing height has $Q_{\rm s}=0.5$, 1, 1.5, and 2
for $Q_{\rm g}=1$. The sequence of red curves with decreasing height
has $Q_{\rm g}=0.5$, 1, and 1.5, with $Q_{\rm s}=1$. The overlapping
case has a dashed blue and red curve. These two sequences are slightly
different. As $Q_{\rm s}$ increases, the peak in the dispersion
relation at $q\sim1-2$ gets less prominent and the dominant part of the
curve shifts to the right. All that is left to drive the combined
instability at low $Q_{\rm s}$ is the dissipation in the gas, which
extends the instability to high $q$. On the other hand, as $Q_{\rm g}$
increases, the peak at $q\sim1-2$ remains and even gets relatively
stronger as the stellar instability begins to dominate. In all cases,
there is a tail of unstable growth at high $q$ from turbulent
dissipation.

The effect of the dissipation term $\delta$ is shown as a sequence of
blue curves in the right-hand panel of Figure
\ref{rafikov_dispersion_sg}, and the effect of the velocity dispersion
ratio $S=\sigma_{\rm g}/\sigma_{\rm s}$ is shown as a sequence of red
curves. For all of these, $Q_{\rm s}=1$ and $Q_{\rm g}=1$. In addition,
${\hat v}_{\rm rot}={\hat v}_{\max}=8$, $\alpha=0.4$, and $D=1$ (which
enter into the scale heights and do not matter much).  For the blue
curves, $\delta=0$, 0.1, 0.3, 0.5, and 1 with $S=0.5$, going from
bottom to top, and for the red curves $S=0.1$, 0.2, 0.3, 0.5, and 1
with $\delta=0.5$ going from right to left. The overlapping case has a
dashed blue and red curve. The effect of increasing dissipation is to
increase the growth rate at high $q$, leaving all else about the same.
The effect of increasing the gas velocity dispersion relative to the
stars is to decrease the dimensionless wavenumber, i.e., make the
unstable wavelength longer. This change is because of the larger
gaseous scale height as $S$ increases. Along this sequence of $S$, the
dimensionless scale heights are $({\hat H}_{\rm g},{\hat H}_{\rm
s})=(0.052, 0.81),$ (0.10, 0.71), (0.15, 0.65), (0.24, 0.56), and
(0.49, 0.49), respectively.  For the entire sequence of $\delta$ in the
blue curves, $({\hat H}_{\rm g},{\hat H}_{\rm s})$ is the same, $(0.24,
0.56)$.

Low $S$ also makes the gas+star system stable at low $q$ in the
right-hand panel of Figure \ref{rafikov_dispersion_sg}. This is because
the gas self-gravity term is small for small $S$ at fixed $Q_{\rm g}$.
Note that the ratio $qS/Q_{\rm g}$ appears in the first term of
equation (\ref{dr2}) but $S$ does not appear in the second term. As $S$
decreases, the first term becomes smaller compare to the second. For
$Q_{\rm s}=1$ in this Figure \ref{rafikov_dispersion_sg}, the second
term alone, from stars, is not enough to drive an instability when
thickness effects are included. When $q$ is small and $S$ is large, the
first term, from gas, adds enough forcing to make the system unstable.
But when $q$ is small and $S$ is small, the gas does not contribute
enough. Another way to see this is that the ratio $qS/Q_{\rm g}$ equals
$\pi G \Sigma_{\rm g,0}k/\kappa^2$ in physical terms. If $S$ decreases
at constant $Q_{\rm s}$ and $Q_{\rm g}$, then $\Sigma_{\rm
g,0}/\Sigma_{\rm s,0}$ decreases, so the gas becomes less
self-gravitating. (Usually one considers that decreasing $S$
corresponds to colder gas and that seems like it would be more
unstable, but that is only for a constant $\Sigma_{\rm g,0}$ and not
for a constant $Q_{\rm g}$.)

\subsection{Ratio of Perturbed Surface Densities}\label{ratio}

The relative contributions from perturbed stars and gas at the onset of
the instability may be determined from the ratio of perturbed surface
densities using equation (\ref{eq:ss}), which involves the quantity
$Y_{\rm s}$. This ratio depends on the wavenumber and growth rate and
so depends on the solution to the dispersion relation. In dimensionless
terms, $Y_{\rm s}$ equals the second term on the right of equation
(\ref{dr2}).

Figure \ref{rafikov_dispersion_sssg3} (top) shows $\Sigma_{\rm s}/
\Sigma_{\rm g}$ versus dimensionless wavenumber $q$ for all of the
gas+star dispersion relations shown in Figure
\ref{rafikov_dispersion_sg}, using the same order of panels, curve
colors and curve types (the previous curves for single component
solutions are not plotted in Fig. \ref{rafikov_dispersion_sssg3}). The
bottom panels show the surface density ratio normalized to its initial
ratio. Generally, the stellar contribution to the instability is
proportional to the initial star-to-gas ratio. It is equal to this
ratio for $q\leq1$, and less than it for $q>1$.

For the fastest growing wavenumber, which is usually larger than
$q\sim1$ (Fig. \ref{rafikov_dispersion_sg}), the ratio of stars to gas
in the perturbation can be between 0.5 and 1 times the initial ratio.
Figure \ref{rafikov_his} shows histograms of the normalized ratio of
perturbed stars to gas at the wavenumber of peak growth. Each histogram
is taken from a series of 800 runs with $Q_{\rm s}=0.2,$ 0.4, 0.6, up
to 4, and $Q_{\rm g}=0.1$, 0.2, 0.2, up to 4, and for the cases in this
range where there is an instability and where the dimensionless
wavenumber of peak growth is less than 40.  The dissipation parameter
$\delta$ increases from top to bottom in the figure. With more
dissipation, the peaks in the histograms of the left-hand panels, which
are for $S=1$, shift to the left, which means there is an increasing
extension of $\Sigma_{\rm s}\Sigma_{\rm g,0}/ \Sigma_{\rm g}\Sigma_{\rm
s,0}$ to very low values. These changes mean that the contribution from
gas is becoming more important as turbulent dissipation increases.

The right-hand side of Figure \ref{rafikov_his} is for lower relative
gas velocity dispersion, $S=0.2$. Aside from an increase in nearly
pure-gas instabilities at $\Sigma_{\rm s}\Sigma_{\rm g,0}/ \Sigma_{\rm
g}\Sigma_{\rm s,0}\sim0$ with increasing $\delta$, there is not much
change in the position of the peak. The lower cutoff to the main peak
at $\Sigma_{\rm s}\Sigma_{\rm g,0}/ \Sigma_{\rm g}\Sigma_{\rm
s,0}\sim0.4$ means that gas contributes less to the instability when
$S$ is low than when $S$ is high, even when there is dissipation. The
reason for this is that when $S$ is low, the first term on the right in
equation (\ref{dr2}) is low at fixed $Q_{\rm g}$.

To derive the absolute ratio of perturbed stars to gas, we recall that
the unperturbed ratio is $\Sigma_{\rm s,0}/\Sigma_{\rm g,0}=Q_{\rm
g}/SQ_{\rm s}$.  Then the absolute ratio is $Q_{\rm g}/SQ_{\rm s}$
times the relative ratio in Figure \ref{rafikov_his}.  For $S=0.2$,
Figure \ref{rafikov_his} has a peak at a relative ratio of $\sim0.6$,
and so the absolute ratio of perturbation strengths is
\begin{equation}
{{\Sigma_{\rm s}}\over{\Sigma_{\rm g}}}={{Q_{\rm g}}\over{ Q_{\rm
s}}}\times {0.6\over S} \sim 3{{Q_{\rm g}}\over{ Q_{\rm s}}}.
\label{ss}
\end{equation}
For $S=1$ and $\delta=0$, Figure \ref{rafikov_his} has a peak near 0.9,
so $\Sigma_{\rm s}/\Sigma_{\rm g}=\sim 0.9 Q_{\rm g}/Q_{\rm s}$.  For
$S=1$ and $\delta=1$, the peak is near 0.6, so $\Sigma_{\rm
s}/\Sigma_{\rm g}\sim 0.6 Q_{\rm g}/Q_{\rm s}$.  Generally $Q_{\rm
g}<Q_{\rm s}$, so the perturbed mass in stars is less than or
comparable to the perturbed mass in gas for a wide range of cases. When
the $Q$ values are large and the main instability involves strong gas
dissipation, the normalized density ratios in Figure \ref{rafikov_his}
are much lower, $\sim0-0.2$, and then the relative contribution from
stars is less.

A contribution from stars that is comparable to the contribution from
gas in an instability, $\Sigma_{\rm s}/\Sigma_{\rm g}\sim 1$, implies
that spiral arms which form from the instability contain old disk
stars.  This is obvious and generally acknowledged for spiral arms.
However, the same high fraction of disk stars could also be present in
instabilities at the dominant wavenumber that make large globular
clouds, as in dwarf galaxies with low shear or in highly unstable
clumpy galaxies in the early universe \citep{elmegreen09}.  The oldest
disk stars would be less abundant in these clouds than the youngest
disk stars because the stellar velocity dispersion increases with age.
Still, stars of intermediate and young age that did not form in the
clouds could have an excess abundance compared to the surrounding disk.
According to equation (\ref{ss}), this excess in disk stars
approximately equals the excess in gas multiplied by the ratio of
$Q_{\rm g}$ to $Q_{\rm s}$.

\subsection{Instability Thresholds}\label{thresh}

The stability threshold is given by the dispersion relation at a growth
rate of $s=0$.  In the absence of dissipation ($\delta=0$) and
thickness effects (${\hat H}=0$), equation (\ref{dr2}) is a sum of two
terms, $2qS/\left(Q_{\rm g}\left[1+q^2S^2\right]\right)$ and
$2q/\left(Q_{\rm s}\left[1+q^2\right]\right)$, considering that
equation (\ref{eq:approx2}) is a good approximation. Each term has a
maximum over variations in $q$ equal to 1/Q. This is how \cite{wang94}
derived an approximate instability threshold for the two-fluid
instability:
\begin{equation}
{1\over{Q_{\rm tot,WS}}}={{1}\over{Q_{\rm g}}}+{1\over{Q_{\rm s}}}
>1. \label{qws}\end{equation}
This threshold has been widely used and is more convenient than the
exact $Q$ threshold for these conditions that requires solving a cubic
equation \citep{e95}.  Extrapolation of this method with thickness
corrections might suggest that instability of a more realistic thick
disk requires
\begin{equation}
{1\over{Q_{\rm tot,WSH}}}={{1}\over{Q_{\rm g}(1+q{\hat H}_{\rm
g})}}+{1\over{Q_{\rm s}(1+q{\hat H}_{\rm s})}}
>1. \label{qwsh}\end{equation}

Recently, \cite{romeo11} proposed a modified form of the Wang \& Silk
equation that is more accurate in the standard case ($\delta=0$),
\begin{eqnarray}
{1\over{Q_{\rm tot,RW}}}={{1}\over{Q_{\rm g}}}+{W\over{Q_{\rm
s}}}>1\;\;{\rm if}\;\; Q_{\rm s}>Q_{\rm g}\\\equiv{{W}\over{Q_{\rm
g}}}+{1\over{Q_{\rm s}}}>1\;\;{\rm if}\;\; Q_{\rm s}<Q_{\rm g}.
\label{qrw}\end{eqnarray} The weighting function is
\begin{equation}
W(S)={{2S}\over{1+S^2}},\end{equation} where $S$ is the ratio of
velocity dispersions, $\sigma_{\rm g}/\sigma_{\rm s}$.  The weighting
function allows the gas term in equation (\ref{dr2}) to contain $S$.
With thickness effects,
\begin{eqnarray}
{1\over{Q_{\rm tot,RWH}}}= {{1}\over{Q_{\rm g}(1+q{\hat H}_{\rm
g})}}+{W\over{Q_{\rm s}(1+q{\hat H}_{\rm s})}}>1\;\;{\rm if}\;\; Q_{\rm
s}>Q_{\rm g}\\\equiv{{W}\over{Q_{\rm g}}(1+q{\hat H}_{\rm
g})}+{1\over{Q_{\rm s}(1+q{\hat H}_{\rm s})}}>1\;\;{\rm if}\;\; Q_{\rm
s}<Q_{\rm g}. \label{qrwh}\end{eqnarray} We recall here for convenience
that the dimensionless quantity $q{\hat H}$ is identical to the
physical quantity $kH$. We test these approximations by finding their
values at the limit where $s=0$ in equation (\ref{dr2}).  We also test
how well they work when gas dissipation is included.

Figure \ref{rafikov_s_vs_qgas} shows numerous plots of peak growth rate
versus $1/Q_{\rm tot}$. Each panel contains 20 blue curves and 20 red
curves. The 20 curves correspond to values of $Q_{\rm s}=0.2$, 0.4,
...4, and, within each curve, $Q_{\rm g}$ varies from 0.1 to 4 in steps
of 0.1. Thus each curve is a sequence of decreasing peak growth rates
for increasing $Q_{\rm g}$. The peak growth rate is evaluated only up
to a wavenumber of $q=40$.

The left-hand panels plot the Wang \& Silk parameter on the abscissa,
with the blue curves for equation (\ref{qws}) and the red curves for
equation (\ref{qwsh}). The middle column plots the Romeo \& Wiegert
parameter on the abscissa, again using blue to indicate no thickness
terms in that parameter. These two columns use a ratio of dispersions
$S=0.2$, which is typical for spiral galaxies. The right-hand panels
consider $S=1$, for which the Wang \& Silk parameter is the same as the
Romeo \& Wiegert parameter.

Thickness effects are considered in the dispersion relation solutions
for all of the curves in the top four rows (even when these effects are
ignored in the expression for $1/Q_{\rm tot}$). What differs in these
top four rows is the value of the dissipation parameter, $\delta$. The
bottom row solves the dispersion relation (\ref{dr2}) without
dissipation and without thickness terms, taking ${\hat H}_{\rm s}={\hat
H}_{\rm g}=0$; the red and blue curves are the same then.

In all panels, the threshold of stability occurs where the curves
intersect the x-axis, at $s_{\rm peak}=0$. Evidently, the stability
thresholds given by equations (\ref{qws})-(\ref{qrw}) work well in the
situations for which they were designed. For an infinitesimally thin
disk with no dissipation and a velocity dispersion ratio of $S=1$
(bottom right panel), the Wang \& Silk formula works perfectly: all the
curves intercept the $1/Q_{\rm tot,WS}=1$ point at $s_{\rm peak}=0$. In
this limit, the two fluids act as a single fluid because they have the
same velocity dispersion. When $S=0.2$ at zero thickness and no
dissipation (lower left panel), the stability threshold is $1/Q_{\rm
tot,WS}=1.28$ on average for all intercepts. The Romeo \& Wiegert
relation works well for both $S$ values in the $\delta={\hat H}=0$
limits (i.e., in the bottom middle and bottom right panels).

When thickness effects are considered in equation (\ref{dr2}), but
still without dissipation (top panels), both the Wang \& Silk parameter
(red curves: top left and top right panels) and the Romeo \& Wiegert
parameter (red curves: top middle and top right panels) work well in
the sense that they intercept the $x$-axis at $1/Q_{\rm tot}\sim1$.
Thus equations (\ref{qwsh}) and (\ref{qrwh}) are both reasonable
approximations to the stability limit on average.

Dissipation lowers $1/Q_{\rm tot}$ in all cases. Both the red and blue
curves in Figure \ref{rafikov_s_vs_qgas} shift to the left in a
sequence of increasing $\delta$ going from the top row to the fourth
row. With zero thickness, disks are always unstable for large enough
wavenumber when dissipation is included (see discussion after equation
\ref{dr2}), but for finite thickness, as in the top four panels of
Figure \ref{rafikov_s_vs_qgas}, a disk eventually become stable when
$1/Q_{\rm tot}$ is small enough (see discussion after equation
\ref{qgt}).  The blue curves dive into the abscissa at low $1/Q_{\rm
tot,WS}$ and $Q_{\rm tot,RW}$ because of this disk stability. The red
curves approach the abscissa asymptotically as $1/Q_{\rm tot,WSH}$ and
$1/Q_{\rm tot,RWH}$ go to zero because $q$ at peak growth gets very
large when there is dissipation. Then $(1+q{\hat H})$ pulls down the
$1/Q_{\rm tot}$ parameters at large $Q$.

The irregular excursions in the red curves of Figure
\ref{rafikov_s_vs_qgas} are because of a rapidly decreasing wavenumber
at peak growth for increasing $Q$ at very low $Q$. Recall that for a
one-component instability, $k_{\rm peak}=\pi G \Sigma/\sigma^2$, or
$q_{\rm peak}=\left(QS\right)^{-1}$.  This large $q$ at small $Q$
enters into the $(1+q{\hat H})$ factor, and, this, along with slight
variations in ${\hat H}$, makes the excursions.

Figure \ref{rafikov_s_vs_qgas_zeros} shows the average x-axis values
where the growth rates in Figure \ref{rafikov_s_vs_qgas} are between
0.03 and 0.3. This is a reasonable range to indicate either absolute
stability or slow growth in the cases of asymptotic approach to the
abscissa. Blue curves are again for $Q_{\rm tot}$ parameters that
ignore thickness and red are for those that include thickness (where
all of the dispersion relation solutions include thickness). The open
square at $\delta=0$ is the Wang \& Silk equation (\ref{qws}) for
$S=0.2$ and ${\hat H}=0$ in the dispersion relation; the x-symbol is
the Romeo \& Wiegert result for $S=0.2$ and ${\hat H}=0$, and the open
triangle is for both Wang \& Silk and Romeo \& Wiegert at $S$=1 with
${\hat H}=0$. Figure \ref{rafikov_s_vs_qgas_zeros} confirms the result

$1/Q_{\rm tot}$ parameters decreases with increasing dissipation: for
$\delta$ in the range from $0.4-1$, $Q_{\rm tot,WSH}$ and $Q_{\rm
tot,RWH}$ should be less than $\sim1.5-2$ for instability when $S=1$,
and they should be less than $\sim3-4$ for instability when $S=0.2$. If
thickness effects are ignored in the stability condition (blue curves),
then for this same range of $\delta$, $Q_{\rm tot,WS}=Q_{\rm
tot,RW}<0.78-0.84$ for instability when $S=1$, and $Q_{\rm
tot,WS}<0.70-0.74$ and $Q_{\rm tot,RW}<0.89-0.95$ for instability when
$S=0.2$.

These results indicate that unknown dissipation partly compensates for
unknown thickness in the stability condition. That is, $Q_{\rm tot}$
conditions that are evaluated incorrectly without including the
thickness terms (blue curves in Figure \ref{rafikov_s_vs_qgas_zeros})
have a threshold for instability that is still close to 1 when
dissipation is important. This compensation has been evident for a long
time in simpler derivations where one uses the adiabatic expression for
pressure, $P\propto \rho^\gamma$, instead of the perturbed energy
equation (\ref{eq:p2}).  Denser regions are cooler so $\gamma<1$. The
compilation in \cite{myers78} suggests that $\gamma\sim0.25$ at
interstellar densities between 0.1 cm$^{-3}$ and 100 cm$^{-3}$. This
index appears in $Q_{\rm g}$ as $\gamma^{1/2}$ multiplying the velocity
dispersion, $\sigma_{\rm g}$. Thus a version of $Q_{\rm g}$ with
adiabatic cooling and thickness corrections is $\gamma^{1/2}\sigma_{\rm
g}\kappa (1+kH)/\pi G \Sigma_{\rm g}$. Since $kH\sim1$ usually,
$\gamma^{1/2}(1+kH)\sim1$, indicating that cooling effects offset
thickness effects even in a simple derivation.

\section{Azimuthal Instabilities}\label{sect:azim}

The main result of the previous section is that dissipation promotes
gravitational instabilities at high wavenumbers even when a gas+star
disk is stable by conventional criteria.  Thickness effects ultimately
stabilize the disk for high $Q_{\rm g}$ and $Q_{\rm s}$. In the most
extreme reduction of this result to a single stability parameter
$Q_{\rm tot}$, we find that both the \cite{wang94} and the
\cite{romeo11} definitions are reasonably good provided the threshold
is raised by a factor of 2 to 3, depending on the relative dissipation
rate.  For example, if significant energy dissipation in the gas occurs
in 2 crossing times over a perturbed wavelength ($\delta=0.5$), then
$Q_{\rm tot}$ including thickness effects has to exceed about 2 for
stability, instead of the usual value of 1.  This is for radial
instabilities.

A higher threshold is also likely for azimuthal instabilities of the
swing-amplifier type \citep{toomre81}.  There is a second instability
parameter in that case., i.e., $X$ in the notation of \cite{julian66}
and $J$ in the notation of \cite{lau78}.

The $X$ parameter is defined as
\begin{equation}
X={{\lambda_{\rm t}}\over{\lambda_{\rm crit}}}
\end{equation}
where $\lambda_{\rm t}$ is the wavelength of the perturbation in the
azimuthal direction and $\lambda_{\rm crit}=2\pi/k_{\rm crit}$ for
$k_{\rm crit}=\kappa^2/(2\pi G \Sigma)$.  The wavelength in the
azimuthal direction is $2\pi R/m$ for radius $R$ and number $m$ of
perturbations in the azimuthal direction. Thus
\begin{equation}
X={{R\kappa^2}\over{2\pi G \Sigma m}}={{{\hat R}Q}\over {2m}} = {Q\over
{2q_{\rm t}}}
\end{equation}
where quantities have been normalized in the same way as before: ${\hat
R}=R\kappa/\sigma$ and $q_{\rm t}=k_{\rm t}\sigma/\kappa$, where
$k_{\rm t}=2\pi/\lambda_{\rm t}$.

The $J$ parameter is about the same:
\begin{equation} J=\left(m\over R\right)\left({{2\pi G\Sigma}\over
{\kappa^2}}\right)\left({{2\Omega}\over{\kappa}}\right) \left(\mid{{
d\ln\Omega}\over{d\ln R}}\mid\right)^{1/2}={1\over{X}}
\left({{2\Omega}\over{\kappa}}\right) \left(\mid{{
d\ln\Omega}\over{d\ln R}}\mid\right)^{1/2}.
\end{equation}
The terms in parentheses on the right are of order unity for angular
rotation rate $\Omega(R)$. For a flat rotation curve, their product is
1.4.

\cite{toomre64} noted that the azimuthal instability in a rotating
galaxy has the same size limit as the radial instability, i.e.,
$\lambda_{\rm t}<\lambda_{\rm crit}$, which is the requirement that
gravity overcomes the centrifugal force that arises when condensing
material spins up by the Coriolis force. This means that instability is
favored by small $X$ or large $J$.

\cite{lau78} and \cite{bertin89} introduced an effective
two-dimensional stability parameter
\begin{equation}
{1\over{Q_{\rm eff}^2}}={1\over{Q^2}}+{{J^2Q^4}\over 4}.
\end{equation}
The second term is
\begin{equation}
{{J^2Q^4}\over 4}=\left(q_{\rm t}Q\right)^2
\left({{2\Omega}\over{\kappa}}\right)^2 \left(\mid{{
d\ln\Omega}\over{d\ln R}}\mid\right)\sim2\left(q_{\rm t}Q\right)^2
\end{equation}
where the factor 2 is for a flat rotation curve.

The azimuthal instability is stronger if the azimuthal wavenumber
$q_{\rm t}$ is larger. There is no absolute stability threshold if
arbitrarily large $q_{\rm t}$ (or $m$) are allowed, because $Q_{\rm
eff}$ can always be made less than the threshold value of 1 by large
enough $q_{\rm t}$.  However, at high $q_{\rm t}$ there is also
pressure stabilization for an adiabatic gas undergoing azimuthal
instabilities. This is the usual Jeans condition.  Without an adiabatic
gas, this Jeans condition goes away if turbulent dissipation is fast
enough, as shown in previous sections.

Transient growth of shearing spiral waves is still possible. If the
dissipation rate is faster than the shear rate, then transient growth
can lead to runaway growth. What usually stabilizes the swing amplifier
at late times is high pressure at the high wavenumbers that are reached
by the swept-back, tightly-wound wavelets.  With turbulent dissipation,
this gas pressure becomes small. Stabilization at late times then comes
from thickness effects rather than pressure. That is, thickness dilutes
the gravitational force at high wavenumbers by the factor
$(1+kH)^{-1}$.

\section{Discussion}\label{disc}

\subsection{Pervasive Instabilities}\label{sect:perv}

Turbulent dissipation, viscosity, and magnetic fields destabilize
galaxy disks. All three either decrease or remove the threshold column
density for instability. Dissipation does this by removing the
stabilization from turbulent pressure, as shown here, while magnetic
fields and viscosity remove the stabilization from Coriolis forces, as
shown by \cite{e87} and \cite{kim02} for the magnetic case, and by
\cite{gammie96} for the viscous case. All three processes rely on gas.
The effects of dissipation discussed here arise if the gas is not
isothermal or simply adiabatic, but satisfies an energy equation with
excess dissipation over heating in a perturbed region. The particular
form for the dissipation rate assumed here, i.e. proportional to the
initial crossing rate in a perturbation, $\delta\sigma_{\rm g}k$, is
appropriate for supersonic turbulence in the general interstellar
medium \citep{maclow98,stone98}. This result differs qualitatively from
the usual result that cooling makes a gas disk more unstable by
lowering $Q_{\rm g}$. It also differs from the result that adiabatic
gas corresponds to greater instability when the adiabatic index
$\gamma$ is lower.

Our primary concern is the stability of galactic gas. This situation
differs from the rotating gas disks studied by \cite{gammie01},
\cite{rice03}, \cite{meija05}, and others in which gravitational
fragmentation occurs if the dissipation time is less than about half
the rotation time. These authors assume a constant cooling time
$\tau_{\rm c}$ for an energy loss rate $U/\tau_{\rm c}$ with internal
energy density $U$. This would take the place of our term
$\Gamma-\Lambda$ in equation (\ref{eq:p}) and therefore differ from our
assumed rate, $\delta\sigma_gkP/(\gamma-1)$, in equation
(\ref{eqheat}). In our equations, the perturbed cooling time is not
constant but scales with the size of the region, $\tau_{\rm c}\propto
1/k$.  The other studies consider instabilities in optically thick,
subsonic accretion disks around protostars or active galactic nuclei.
In this environment, dissipation is difficult because the radiation
cannot easily escape. Dissipation may be aided by convection in the
lower layers \citep{boss04}, or it may be prevented by slow radiative
losses at the disk surface \citep{cai10}. Galactic gas, in contrast, is
optically thin and supersonic. Dissipation is by random shocks and
energy loss is by unblocked radiation from the shock fronts. In our
case, small-scale instabilities are possible in the rapidly cooled
shocked regions, even if they are smaller than the Jeans length at the
initial velocity dispersion.

The present result implies that galaxy disks may be more unstable than
previously determined from models or computer simulations with
isothermal or adiabatic gas. The effective $Q$ threshold for gas+star
radial instabilities increases to $\sim2$ if turbulent dissipation
takes a few crossing times. This increased $Q_{\rm tot}$ value is
comparable to what \cite{leroy08} found for most regions of their
galaxies, which implies that galactic disks are only marginally stable.

Spiral arms are examples of gravitational instabilities that depend on
$Q_{\rm tot}$. Multiple arm galaxies have what \cite{tk91} called
spiral chaos, with gas+star spiral features of all sizes forming
everywhere in the disk and continuously replacing old ones that get
sheared away. The gaseous parts of disks can have independent
instabilities that make spiral arms in a flocculent pattern of star
formation \citep[e.g.,][]{e11}. Multiple arm and flocculent structure
are instabilities covering a wide range of scales. They should pump
turbulent energy into the gas \citep[e.g.,][]{bournaud10} and heat the
stars \citep{carlberg85,sellwood11}. Grand design spirals may be
triggered by interactions or bars \citep{kormendy79} but require a
marginally stable disk to amplify these perturbations to large
arm/interarm contrasts.

Power spectra of optical light in galaxies, including grand-design
galaxies like M81, are continuous power-laws suggesting pervasive
instabilities and a cascade of power over a wide range of scales
\citep{e03,e06}. There is no dominance of structure at the
fastest-growing wavelength of an instability. Such lengths are a
feature of the linear regime and may appear only in young regions that
start fairly smooth. Examples might be expanding shells or spiral arm
shocks, which sometimes show a characteristic clump scale
\citep{deharveng10, e06b}.

\subsection{Feedback Regulation of Disk Stability}\label{sect:feed}

The observation by \cite{leroy08} that $Q_{\rm tot}\sim2$ for most
galaxy disks in their survey implies some regulatory mechanism.
Regulation is usually considered to involve star formation so that low
$Q_{\rm tot}$ triggers more star formation, which increases
$\sigma_{\rm g}$ and $Q_{\rm tot}$ \citep{goldreich65}. High $Q_{\rm
tot}$ triggers less star formation, causing $\sigma_{\rm g}$ to
decrease by passive cooling, and lowering $Q_{\rm tot}$ to a more
unstable value.  $Q_{\rm tot}$ also affects the stellar velocity
dispersion because growing instabilities scatter disk stars. The
present results detach $\sigma_{\rm g}$ from $Q_{\rm tot}$ somewhat,
because dissipation also lowers $\sigma_{\rm g}$ during the initial
growth of the instability.  Dissipation and instabilities are not two
separate steps in this model.

The variety of spiral arm strengths in galaxies suggests that $Q_{\rm
tot}$ is not strongly regulated.  The stellar spirals in multiple arm
and flocculent galaxies vary in strength, from 50\% or stronger surface
density modulations in multiple arm galaxies to no perceptible
modulations in some flocculent galaxies.   Of the 197 flocculent
galaxies catalogued by \cite{ee87}, 85\% have no perceptible long-arm
spirals at 2$\mu$ in the 2MASS survey \citep{e11}. This result suggests
that flocculent galaxies have more stable stellar disks than multiple
arm galaxies, which means that not all $Q_{\rm tot}$ values are the
same. Similarly, in a sample of 147 Ohio State University Bright Galaxy
Survey galaxies less inclined than $65^\circ$, the distribution
function of spiral arm torques is broad and featureless, spanning a
factor of $\sim4$ \citep{buta05}. Spiral arm morphology and strength
are the most direct indicators of $Q_{\rm tot}$, because these
structures are what gas+star instabilities make. Star formation does
not have such a clear connection with $Q_{\rm tot}$.

The composite stability conditions shown in Figure
\ref{rafikov_s_vs_qgas} illustrate the conventional notion that
instabilities maintain $Q_{\rm tot}$ at a marginally unstable value by
stirring the stars and gas. If $Q_{\rm tot}$ is small then the
threshold, $s_{\rm peak}$, is large and stirring is active, which
lowers $Q_{\rm tot}$. Feedback is complex for a gas+star fluid,
however, because the stars do not cool.  If star formation is active,
then cool stars can be added continuously to the stellar population,
which is like cooling \citep{carlberg85}, but the same instabilities
also heat the old stars in proportion. Stronger instabilities heat more
old stars and, through cloud formation, produce more cool stars. If
heating beats cooling in this proportion, then there would be no
feedback to make the stars cooler if $Q_{\rm tot}$ is too high. If
cooling from new stars beats heating, then there would be no feedback
to make the stars heat up if $Q_{\rm tot}$ is too low.

Feedback stabilization of $Q_{\rm tot}$ is also complex for a gas+star
fluid because the cool young stars have the same velocity dispersion
and mass as the gas did before the stars formed.  Thus the new, cool
stars do not add any extra instability to the combined disk. They
remove the ability to dissipate energy however, which is a stabilizing
effect.

Persistent heating of old stars by gravitational instabilities requires
more and more cool gas to maintain a marginal $Q_{\rm tot}$ if there is
regulation. This eventually leads to a gas instability that is
decoupled from the star instability, i.e., when $S=\sigma_{\rm
g}/\sigma_{\rm s}<<1$. Then there are two peaks in the dispersion
relation, one for stars at low $q$ and another for gas at high $q$
\citep{jog84,romeo92}. Some flocculent galaxies could be in this
regime.

If the stellar $Q_{\rm s}$ is low, then stellar spirals form readily
and the disk cannot be stabilized by extra turbulence following star
formation. Feedback comes only from stellar heating in the waves. On
the other hand, if $Q_{\rm s}$ is high, then the combined instability
shows up primarily in the gas, producing flocculent spirals and
filaments of star formation.

Galactic gas accretion can modify $Q_{\rm tot}$ by adding cold gas, and
after star formation, by adding slow-moving stars. Following a sudden
influx of gas, $Q_{\rm tot}$ should drop and a period of intense spiral
arm or clump formation and star formation should begin. When the gas
eventually gets converted to stars, $Q_{\rm tot}$ should increase and
the spiral arms weaken. This would be a regulatory process that does
not connect $Q_{\rm tot}$ with star formation directly but still keeps
$Q_{\rm tot}$ at a more-or-less constant value, particularly if the
star formation rate is about equal to the galactic accretion rate on
average.

\subsection{Star Formation}\label{sf}

\subsubsection{Is there a Connection with Galactic-Scale Gravitational
Instabilities?}

The direct role of gravitational instabilities in star formation is
difficult to observe. According to \cite{leroy08}, \cite{bigiel08} and
others, star formation only requires molecular gas, and molecular gas
only requires sufficient pressure, most of which comes from the weight
of the gas layer in the disk. There is nothing about cloud formation or
possible instabilities that induce cloud formation in these conditions.
Only pressure determines the abundance of dense, star-forming clouds.
\cite{krum09} and \cite{ostriker10} reproduce the observed radial
profiles of galaxies using only the constraint that molecular clouds
are shielded by atomic gas in some type of equilibrium. It is difficult
to see how galactic-scale gravitational instabilities can influence
these results as the instabilities operate on the total gas (and stars)
without regard to molecular phase.  If gravitational instabilities
directly triggered star formation, then the primary scaling law should
have a star formation rate that depends on a high power, such as 1.5 or
2, of the total gas column density \citep[e.g.,][]{madore77,li05}, and
this is apparently not observed \citep[however, see][]{yang07,liu11}.
For a review of star formation processes, see \cite{mckee07}.

Perhaps this disconnection between star formation and gravitational
instabilities makes sense in retrospect. If the gas is already highly
molecular, then the conversion of a little extra atomic material into
molecular material by compression in instabilities will not influence
the total molecular mass and star formation rate much, nor will any
rearrangement of the existing molecular clouds by the unstable motions.
Even cloud collisions during an instability would not affect the star
formation rate if the molecular depletion time is as constant as it
appears to be \cite{leroy08}. On the other hand, if the gas is highly
atomic, then instabilities could promote molecule formation if the
density contrast becomes large. However, on the dominant scale of the
gas+star instability, which is $\sigma_{\rm s}/\kappa$ (i.e.,
$q\sim1$), and for the dominant growth rate, $\omega\sim\kappa$ (i.e.,
$s\sim1$), these instabilities would only increase the density by
several exponential growth factors during the available shear time.
This is not enough compression to make molecular clouds. Star formation
has to occur in the non-linear regime of unstable growth, or
independently of this growth, when the gas separates from the stars and
collapses into dense molecular clouds by its own gravity
\citep[e.g.,][]{ko02}. The stars that helped start the instability will
have moved away in their epicycles after this time, with little
long-term excess following the clumped gas. Star formation also occurs
after the instability causes contraction in the third dimension,
considering that giant molecular clouds (GMCs) are smaller than the
scale height. In addition, star formation can occur in pressurized
shells and cometary structures that are unrelated to galactic-scale
instabilities.

The observation of a linear molecular-cloud star formation law, which
suggests a secondary role for galactic-scale gravitational
instabilities, combined with the observation of a weakly-varying
$Q_{\rm tot}$, which suggests feedback-regulation of $Q_{\rm tot}$ by
gravitational instabilities that may trigger star formation, poses an
interesting contradiction. Evidently the role of the these
instabilities in star formation is indirect.

One possibility is that gravitational instabilities replenish the
self-gravitating molecular gas that gets converted into diffuse gas
during the star formation process. This way the molecular gas abundance
remains constant on average, and the star formation rate does not scale
with the cloud formation rate, but only with the current molecular
abundance. An example is provided by the simulations in
\cite{bournaud10} and \cite{hopkins11}, which show gravitational
instabilities that form self-gravitating gas from diffuse gas, and star
formation that disrupts these gravitating clouds, converting a fraction
of their gas back into diffuse form.  The conversion rate of diffuse
gas into self-gravitating gas by galactic-scale instabilities can be
much larger than the conversion rate of self-gravitating gas into stars
by local collapse. A simple model for this is discussed in the next
sub-section.

\subsubsection{A Simple Model of Star Formation with a Role
for Gravitational Instabilities}

A simple model of star formation illustrates how gravitational
instabilities can play an important role in cloud formation, which is
cloud re-assembly in this picture, and yet not trigger star formation
directly or have a direct connection to the star formation rate. In
this model, there are three phases of interstellar matter:
self-gravitating clouds in which stars form and which are generally
molecular, diffuse molecular clouds in which stars do not form because
their density is too low, and diffuse atomic or ionized clouds, which
are also too low in density to form stars. If, in a certain region
their masses are $m_3$, $m_2$, and $m_1$, respectively, then we can
write the equations governing their changes:
\begin{equation}
{{dm_3}\over{dt}}={{m_1+m_2}\over{T_{\rm GI}}}-{{d_3m_3}\over{T_{\rm
burst}}}-{{m_3}\over{T_{consume}}},\label{model1}\end{equation}
\begin{equation}
{{dm_2}\over{dt}}=-{{m_2}\over{T_{\rm GI}}}+{{d_2m_3}\over{T_{\rm
burst}}},\label{model2}\end{equation}
\begin{equation}
{{dm_1}\over{dt}}=-{{m_1}\over{T_{\rm GI}}}+{{d_1m_3}\over{T_{\rm
burst}}}\label{model3}.\end{equation} The timescales and terms in these
equations are as follows.  $T_{\rm GI}=1/\omega$ is the timescale for
gravitational instabilities studied here. For most cases, this is
comparable to the epicyclic time, which means that $s\sim1$ in the
dimensionless notation used above. Then the first equation says that
the self-gravitating mass increases on the timescale of the
gravitational instability with the mass coming from diffuse gas. That
is, gravitational instabilities compress all of the gas into discrete
self-gravitating clouds, but the practical effect of this for star
formation is to move the diffuse gas that lies between existing GMCs
onto those GMCs, and to make new GMCs.

$T_{\rm burst}$ is the duration of a single event of star formation in
any one cloud, such as the lifetime of an active OB association. It
should be something like 10-20 Myr for a $10^5\;M_\odot$ cloud. The
factor $d_3$ is the fraction of a self-gravitating cloud that is
converted into diffuse gas during each burst; it might be 30\% or more,
considering disruption of a cluster-forming core and all of the
associated ionization. It is not likely to be 100\% because dense
clouds are mostly pushed to the side by pressures from ionization and
stellar winds, into shell-like and cometary structures, without
complete disassociation into atoms.  Of this disruption fraction, part
of it presumably stays in molecular form but at too low a density to
form stars immediately. It could be molecular H$_2$ without much CO,
for example. Many local diffuse clouds are in this state
\citep{spitzer75}. \cite{rahman10} discuss observations of the star
formation law with explicit consideration of diffuse molecular gas. It
could also be in the form of GMC envelopes that are exposed to
background starlight and have high ionization fractions and long
magnetic diffusion times (Elmegreen 2007). With this term in equation
(\ref{model1}), the self-gravitating mass decreases because of cloud
disruption on the timescale of the individual burst.

$T_{\rm consume}$ is the consumption time for self-gravitating gas in
general. The consumption time is the total self-gravitating molecular
cloud mass in a galaxy divided by the total star formation rate.  It is
usually measured to be $\sim2$ Gyr for main galaxy disks
\citep{leroy08, bigiel08}.  The star formation rate is this last term,
$m_3/T_{\rm consume}$.

There are also parameters in these equations that indicate the
fractions of the various phases that convert into other phases. The
parameter $d_2$ is the fraction of the gravitating cloud mass that
turns into diffuse molecular gas during a burst, and $d_1$ is the
fraction that turns into diffuse atomic or ionized gas; then
$d_1+d_2=d_3$.

Equations (\ref{model1})-(\ref{model3}) can be solved numerically for
the time dependence of the three phases, but the result of interest
here comes more easily by considering the dominant terms. In equation
(\ref{model1}), the last term represents the conversion of gas into
stars. Because of the low efficiency of star formation, which means the
large value of $T_{\rm consume}$ compared to $T_{\rm burst}$ and
$T_{\rm GI}$, this third term is relatively small. In that case, there
is a quasi-equilibrium between the three phases in which the time
derivatives are small compared to each separate term involving $T_{\rm
GI}$ and $T_{\rm burst}$. In this equilibrium,
\begin{equation}
m_1\sim{{m_3T_{\rm GI}}\over{T_{\rm burst}}}d_1\;\;;\;\;
m_2\sim{{m_3T_{\rm GI}}\over{T_{\rm burst}}}d_2.
\end{equation}
What we would like is a measure of how active the gravitational
instabilities are in rebuilding the clouds that form stars. The
building rate is $(m_1+m_2)/T_{\rm GI}$, because the instability
collects all of the gas and places it in dense clouds, which means it
converts the diffuse gas into self-gravitating gas.  This building rate
should be compared with the star formation rate, which is $m_3/T_{\rm
consume}$.  The ratio is
\begin{equation}
\left({{m_1+m_2}\over{T_{\rm GI}}}\right)\left({{T_{\rm consume}} \over
{m_3}}\right) = {{d_3T_{\rm consume}}\over {T_{\rm burst}}}
\end{equation}
For $d_3\sim0.3$, $T_{\rm consume}\sim1$ Gyr, and $T_{\rm burst}\sim
20$ Myr, the ratio is 15.

The point of this simple model is to show how star formation can happen
in self-gravitating clouds at a fixed rate, independent of
gravitational instabilities, i.e., without direct triggering, and yet
these instabilities are important in maintaining the cloud population
so that this star formation processes can continue. The mass flux into
self-gravitating clouds is 15 times the mass flux into stars in this
example.  In practice, other processes of dense cloud formation, such
as shell formation around superbubbles, will combine with gravitational
instabilities to return diffuse debris back into self-gravitating
clouds.

There are many computer simulations of star formation in galaxy disks
that contain the basic elements discussed here, namely gravitational
instabilities that drive turbulence and dense cloud formation, star
formation in the dense clouds with a certain fixed or density-dependent
rate, and dense cloud disruption
\citep[e.g.,][]{wada07,robertson08,dobbs09,baba09,
gnedin10,bournaud10,murante10,gnedin11,ostriker11,tasker09,tasker11}.
These generally have an equation of state for the gas that includes the
heightened instability from dissipation that is discussed in the
present paper.

\section{Summary}\label{sect:sum}

Gravitational instabilities in a gas+star galaxy disk are enhanced when
gas dissipation is allowed to occur during the growth of the
instability. Then the restoring force from pressure and the velocity
dispersion are reduced during the unstable growth. This has the effect
of removing the minimum wavelength (the Jeans length) from the problem,
and for a thin disk, promoting an unstable condition regardless of $Q$.
For a thick disk, the thickness itself is what limits or prevents
instabilities by diluting self-gravity in the in-plane direction
compared to the Coriolis force.

This paper derived the dispersion relation for small perturbations in a
dissipative gas + stellar disk. The effects of disk thickness were
considered in detail, including a self-consistent evaluation of the
thicknesses of the two components.   Threshold effects for an
appropriately defined parameter $Q_{\rm tot}$ were determined and found
to differ from the usual results determined without explicit inclusion
of dissipation. That is, $Q_{\rm tot}$ has to be a factor of 2 to 3
higher than the usual value of unity before a reasonably thick disk
becomes significantly stable. This change implies that real galaxy
disks are more unstable than previously believed.

The greater instability has implications for spiral waves and cloud
formation, and for feedback control of $Q_{\rm tot}$, which is much
more complicated for a two-component disk than a one-component disk,
considering that an existing stellar component can never cool.  The
instability also has implications for star formation laws, particularly
if they suggest a linear dependence between the star formation rate and
the molecular abundance. Such laws would not seem to have a role for
gravitational instabilities in directly triggering star formation, but
they may still have a role in re-assembling the self-gravitating clouds
that are torn apart by stellar feedback. The mass exchange in this
assembly phase can exceed the mass flow into stars by a factor of 10 or
more, indicating the importance of the instability as a component of
the whole process.

\appendix

\section{Scale Heights}\label{sect:height}

The vertical profiles for perturbed gas and stars in our model are the
same as the vertical profiles for equilibrium gas and stars because all
of the motions are assumed to be parallel to the disk plane. The scale
heights then come from the equations of vertical equilibrium. We follow
\cite{narayan} and write the equation for density of each component
$i=$ $g$ and $s$:
\begin{equation}
{{d^2\rho_{\rm i}}\over {dz^2}}={\rho_{\rm i}\over{\sigma_{\rm i}^2}}
\left(-4\pi G(\rho_{\rm g}+\rho_{\rm s})+{{dK_{\rm DM}}\over {dz}}
\right)+{1\over{\rho_{\rm i}}}\left({{d\rho_{\rm i}} \over
{dz}}\right)^2,\end{equation} where the dark matter acceleration per
unit length is
\begin{equation}
{{dK_{\rm DM}}\over{dz}}=-{{v_{\rm max}^2}\over {r^2}} \left(
1-{{2z^2}\over{r^2}} - {{z^2R_{\rm c}^2}\over {r^2(R_{\rm c}^2+r^2)}} +
\left[{{R_{\rm c}}\over r}\right]\left[ {{3z^2}\over
{r^2}}-1\right]{\rm arctan}\left[r/R_{\rm c}\right] \right)
\label{eq:kdm}
\end{equation}
for a dark matter density distribution
\begin{equation}
\rho_{\rm DM}(r)={{\rho_{\rm DM0}R_{\rm c}^2}\over{R_{\rm c}^2
+r^2}}\equiv {{v_{\rm max}^2}\over {4\pi G(R_{\rm
c}^2+r^2)}},\label{eq:dm}\end{equation} with central density $\rho_{\rm
DM0}$, core radius $R_{\rm c}$, 3D radius
$r=\left(R^2+z^2\right)^{1/2}$, disk radius in cylindrical coordinates
$R$, and extrapolated rotation speed at infinity, $v_{\rm max}$.

The rotation curve enters into our primary variable $\kappa$, which is
used for normalization, so we should be self-consistent in the
definitions of these quantities.  We write the rotation curve in the
midplane as the sum of contributions from the dark matter and the disk,
\begin{equation}
v_{\rm rot}^2(R)=v_{\rm rot,DM}^2(R)+v_{\rm rot,Disk}^2(R)
\end{equation}
with
\begin{equation}
v_{\rm rot,DM}^2(R)={{ G M_{\rm DM}(R)}\over R}\end{equation} and
\begin{equation}
M_{\rm DM}(R)=\int_0^R 4\pi R^2\rho_{\rm DM} dR= 4\pi \rho_{\rm
DM0}R_{\rm c}^3 \left({R\over {R_{\rm c}}}-{\rm arctan}{R\over {R_{\rm
c}}} \right).
\end{equation}
We assume that the disk mass column density profiles are exponential,
$\Sigma_{\rm i,0}= \Sigma_{\rm i,00}e^{-R/R_{\rm D}}$, and have the
same disk scale lengths $R_{\rm D}$ for gas and stars. Then
\begin{equation}
v_{\rm rot,Disk}^2(R)=4\pi G\left(\Sigma_{\rm g,0}+\Sigma_{\rm
s,0}\right)R_{\rm D}y^2 \left(I_0\left[y\right]K_0\left[y\right]
-I_1\left[y\right] K_1\left[y\right]\right)\end{equation} in the
notation of Binney \& Tremaine (2008, eqn. 2.165), with modified Bessel
functions $I$ and $K$, and $y=R/2R_{\rm D}$.

Note that as $R\rightarrow\infty$, $v_{\rm rot}^2\rightarrow 4\pi G
\rho_{\rm DM0} R_{\rm c}^2,$ which is $v_{\rm max}^2$ in equation
(\ref{eq:dm}). Now we write
\begin{equation}\kappa^2=2\left({{v_{\rm rot}}\over R}\right)^2\left(1+{{d\log v_{\rm
rot}}\over {d\log R}}\right)\end{equation} and define $\alpha=d\log
v_{\rm rot}/d\log R$.  This gives a relation between the dimensionless
disk radius ${\hat R}$ and the dimensionless rotation speed,
\begin{equation} {\hat R}^2=2(1+\alpha){\hat v}_{\rm rot}^2 .\end{equation}
The other dimensionless parameters are related also. From equation
(\ref{eq:dm}), the ratio of the dark matter central column density to
the stellar disk central column density is $\pi\rho_{\rm DM0}R_{\rm
c}/\Sigma_{\rm s,00}$. If we let $D=\rho_{\rm DM0}R_{\rm c}/\Sigma_{\rm
s,00}$, then
\begin{equation}
{\hat R}_{\rm c}={{{\hat v}_{\rm max}^2Q_{\rm
s}}\over{4D}}.\end{equation}

In terms of the characteristic dimensions used for section
\ref{sect:eq}, namely $\kappa$ for inverse time and $\sigma_{\rm s}$
for velocity, the vertical equilibrium equations in dimensionless form
become
\begin{equation}
{{d^2{\hat \rho}_{\rm g}}\over {d{\hat z}^2}}={{\hat \rho}_{\rm
g}\over{S^2}} \left(-{2\over{Q_{\rm s}}}({\hat \rho}_{\rm g}+{\hat
\rho}_{\rm s})+{{d{\hat K}_{\rm DM}}\over {d{\hat z}}}
\right)+{1\over{{\hat \rho}_{\rm g}}}\left({{d{\hat \rho}_{\rm g}}
\over {d{\hat z}}}\right)^2. \label{drdz2}\end{equation}
\begin{equation}
{{d^2{\hat \rho}_{\rm s}}\over {d{\hat z}^2}}={\hat \rho}_{\rm g}
\left(-{2\over{Q_{\rm s}}}({\hat \rho}_{\rm g}+{\hat \rho}_{\rm
s})+{{d{\hat K}_{\rm DM}}\over {d{\hat z}}} \right)+{1\over{{\hat
\rho}_{\rm s}}}\left({{d{\hat \rho}_{\rm s}} \over {d{\hat
z}}}\right)^2. \label{drdzs2}\end{equation} Equation (\ref{eq:kdm}) for
$dK_{\rm DM}/dz$ looks essentially the same in dimensionless form with
dimensionless distances like ${\hat r}=r\kappa/\sigma_{\rm s}$
substituted for physical distances, and with the dimensionless rotation
speed ${\hat v}_{\rm rot}=v_{\rm rot}/\sigma_{\rm s}$. In equations
(\ref{drdz2}) and (\ref{drdzs2}), dimensionless density is related to
physical density as ${\hat \rho}=2\pi G\rho Q_{\rm s}/\kappa^2$. Note
that for a single fluid, the solution to equation (\ref{drdz2}) would
be ${\hat \rho}({\hat z})={\hat \rho}({\hat z}=0){\rm sech}^{2}({\hat
z/{\hat H}})$ and the dimensionless scale height would be ${\hat
H}=S\left(Q_{\rm s}/{\hat \rho}[{\hat z}=0]\right)^{1/2}$. The
dimensionless column density would be ${\hat
\Sigma_0}=\int_{-\infty}^{\infty}{\hat \rho}d{\hat z}=2{\hat
\rho}({\hat z}=0){\hat H}$. For a single fluid in physical units,
$H=\sigma^2/\left(\pi G \Sigma_0\right)=\sigma/\left(2\pi G\rho[z=0]
\right)^{1/2}$.

The vertical equilibrium equations may be solved for $\rho_{\rm i}(z)$
and the scale heights defined as $H_{\rm i}=0.5\Sigma_{\rm
i,0}/\rho_{\rm i} (z=0)$. For the dimensionless solutions, the
dimensionless input parameters are $Q_{\rm s}$, $S=\sigma_{\rm
g}/\sigma_{\rm s}$, ${\hat v}_{\rm rot}$, ${\hat v}_{\rm max}$, and the
dark matter-to-disk column density ratio, $D$. Also entering is the
logarithmic gradient of the rotation curve, $\alpha$, which varies with
radius and depends on $D$, and disk-to-halo size ratio, $R_{\rm
D}/R_{\rm c}$.  The unperturbed gas-to-star column density ratio,
$\Sigma_{\rm g,0}/\Sigma_{\rm s,0}$ may be treated either as an
independent variable, or it can be taken equal to $SQ_{\rm s}/Q_{\rm
g}$ if $Q_{\rm g}$ is independently specified.

In practice, we do not need a detailed model of a galaxy just to get an
estimate of the scale heights. To simplify things, representative
values of $\alpha$ were determined from the average over a range of
parameters: $D=(0,2)$, $\Sigma_{\rm g,0}/\Sigma_{\rm s,0}=(0,2)$, and
$R_{\rm D}/R_{\rm c}=(0.2,2)$. The average $\alpha$ for disk position
$R/R_{\rm c}=1$ was found to be 0.4, meaning the rotation curve is on
the rising part there, and the average for $R/R_{\rm c}=4$ is 0.04,
which is on the flat part. We consider these two cases in what follows.
The first case can apply to both the inner regions of massive disk
galaxies and to most of the visible parts of the disks of dwarf
galaxies \citep[e.g.,][]{burstein85}.

Figure \ref{rafikov_jog} shows the profiles of $\rho_{\rm g}(z)$ and
$\rho_{\rm s}(z)$ for several dark matter ratios $D$ and velocity
dispersion ratios $S$. The solid curves are for $\alpha=0.4$ and the
dashed curves for $\alpha=0.04$; they hardly differ from each other.
For all colors and line types, the lower curves are for gas and the
upper curves for stars. When $S=\sigma_{\rm g}/\sigma_{\rm s}=1$ (blue
curves), the gas scale height is the same as the stellar scale height
and the profiles differ only in amplitude by the ratio of column
densities ($\Sigma_{\rm g,0}/\Sigma_{\rm s,0}=0.2$). When $S$ is small,
the gas disk is thin and the dark matter ratio does not affect the gas
scale height (red and green curves for gas are about the same).  The
stellar profile does not depend much on $S$ (blue and green curves for
stars are about the same), but it does depend on the dark matter ratio
in the sense that less dark matter makes the stellar disk thicker (red
curve for stars is higher than the blue or green curves).  In all
cases, $Q_{\rm s}=2$, ${\hat v}_{\rm rot}={\hat v}_{\rm max}=8$, and
$\Sigma_{\rm g,0}/ \Sigma_{\rm s,0}=0.2$.

Figure \ref{rafikov_jogh} shows the dimensionless scale heights versus
the gas fraction in the disk, $\Sigma_{\rm g,0}/\Sigma_{\rm s,0}$, on
the left, versus the velocity dispersion ratio $S$ in the middle, and
versus $Q_{\rm s}$ on the right. These are the parameters that cause
the scale heights to vary most. In all cases, ${\hat v}_{\rm rot}={\hat
v}_{\rm max}=8$, and $\alpha=0.4$. Equations (\ref{drdz2}) and
(\ref{drdzs2}) show why the gas scale height increases with $S$ and
both scale heights increase with $Q_{\rm s}$ for this normalization. In
the left and middle panels, $Q_{\rm s}=2$. Both dimensionless scale
heights decrease with increasing $\Sigma_{\rm g,0}/\Sigma_{\rm s,0}$
because then the gas density increases at fixed $\Sigma_{\rm s,0}$ and
this pulls in the star layer. Analogous results for $\alpha=0.04$ and
for a wide range of ${\hat v}_{\rm rot}$ are nearly identical to those
in Figure \ref{rafikov_jogh}.

Representative values for normal spiral galaxy disks in the local
universe, where $S\sim0.2-0.4$, and $\Sigma_{\rm g,0}/\Sigma_{\rm
s,0}\sim0.2$, are ${\hat H}_{\rm g}\sim0.3$ and ${\hat H}_{\rm
s}\sim1.3$.  For clumpy galaxies at high redshift and local dwarfs,
$S\sim0.8-1$, and $\Sigma_{\rm g,0}/\Sigma_{\rm s,0}\sim1$, so ${\hat
H}_{\rm g}\sim0.7$ and ${\hat H}_{\rm s}\sim0.9$. The ratio of the
rotation speed to the velocity dispersion of the stars, the slope of
the rotation curve, and the ratio of the dark matter to the disk mass
do not influence these normalized scale heights significantly.

\section{Thickness Correction Factors}\label{correct}

Equation (\ref{eq:grav}) contains an approximation for disk thickness
that can be evaluated using the full solutions for the vertical density
profiles just calculated. If $\rho(z)=\rho_{\rm g}(z)+\rho_{\rm s}(z)$
is the vertical distribution of total density in a plane-parallel
layer, and if this total density has a $\cos (kx)$ dependence in the
direction of the perturbation, as assumed above, then the acceleration
measured at position $x_0$ in the midplane in this direction is
\begin{equation}
g(x_0)=2G\int_{-\infty}^{\infty} dz \int_{-\infty}^{\infty} dx
{{\rho(z) \cos(kx)(x-x_0)}\over
{{z^2}+(x-x_0)^2}}.\label{gx0}\end{equation} This comes from the
acceleration $2G\mu/r$ toward a line of mass with mass per unit length
$\mu=\rho dx dz$ measured at a distance $r=(z^2+[x-x_0]^2)^{1/2}$.
Converting equation (\ref{gx0}) to dimensionless units as above,
separating the gaseous and stellar parts of $\rho$, setting
$g=-d\phi/dx$, and using an equation like (\ref{eq:grav}) with
arbitrary correction factors $C$ instead of $(1+kH)^{-1}$, we get
\begin{equation}
{{C_g\Sigma_{\rm g}}\over {\Sigma_{\rm s}}} + C_{\rm s}= {{-1}\over
{\pi\sin (qx_0)}}\int_0^{\infty} d{\hat z}\int_{-\infty}^{\infty}
d{\hat x}{{{\hat \rho}({\hat z})\cos q{\hat x}({\hat x}-{\hat
x}_0)}\over {{\hat z}^2+({\hat x}-{\hat
x}_0)^2}}\label{eq:cc}.\end{equation} Note that the cosine term inside
the integral gets converted to a sine term after integration, and this
phase dependence is divided out by the sine in the denominator. This is
consistent with the fact that a cosine variation in density produces a
sine variation in gravitational acceleration, and this produces a
cosine variation in the potential, which is used in equation
(\ref{eq:grav}).  The integral was evaluated for a variety of cases
over the range $q=(0,20)$ and the coefficients $C_{\rm i}=(1+q{\hat
H_{\rm i}})^{-1}$ were found to be good approximations.

Figure \ref{rafikov_grav2} shows the ratio of the right hand side of
equation (\ref{eq:cc}) to the left hand side using $C_{\rm i}=(1+q{\hat
H}_{\rm i})^{-1}$, plotted as a function of dimensionless wavenumber
$q$. This ratio is the correction factor that should be applied to the
right-hand side of equation (\ref{eq:grav}), or to the column
densities, to make the thickness correction more accurate when the
approximation $(1+kH)^{-1}$ is used. Alternatively, the inverse of this
correction factor can be applied to the stability parameter $Q$ to get
a more accurate stability parameter. Fortunately, the correction factor
is small, less than $15$\% in all cases, and only that large near
$kH\sim1-4$. In the range $kH\sim1-10$, the average correction in
Figure \ref{rafikov_grav2} is 12\%, which means that the true
acceleration that drives the instability is higher than the estimate
using the potential of equation (\ref{eq:grav}) by an average of
$\sim12$\%.  The instability growth rate scales as the square root of
the acceleration per unit length, so the likely correction to the
growth rate is an increase by $\sim6$\%. Considering the other
approximations used here for interstellar and stellar dynamics, this
correction to the growth rate is not important. We therefore evaluate
the dispersion relation using the original Vandervoort approximation,
as given by equation (\ref{eq:grav}).

\clearpage
\begin{figure}
\epsscale{.9} \plotone{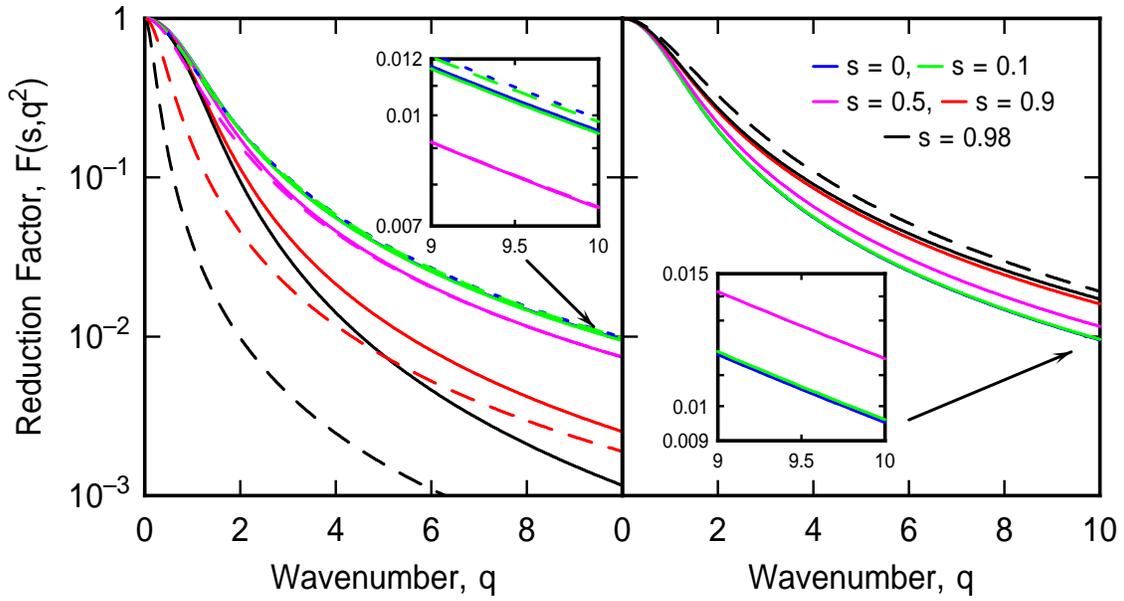} \caption{The reduction factor ${\cal F}$
for gravitational force is plotted versus the dimensionless wavenumber,
$q$, for five different dimensionless rates, $s$, indicated by color.
Oscillating solutions are shown on the left and growing solutions are
shown on the right. Dashed lines are approximate solutions from
equations (6) and (13). }\label{rafikov_reduction}
\end{figure}

\begin{figure}
\epsscale{.9} \plotone{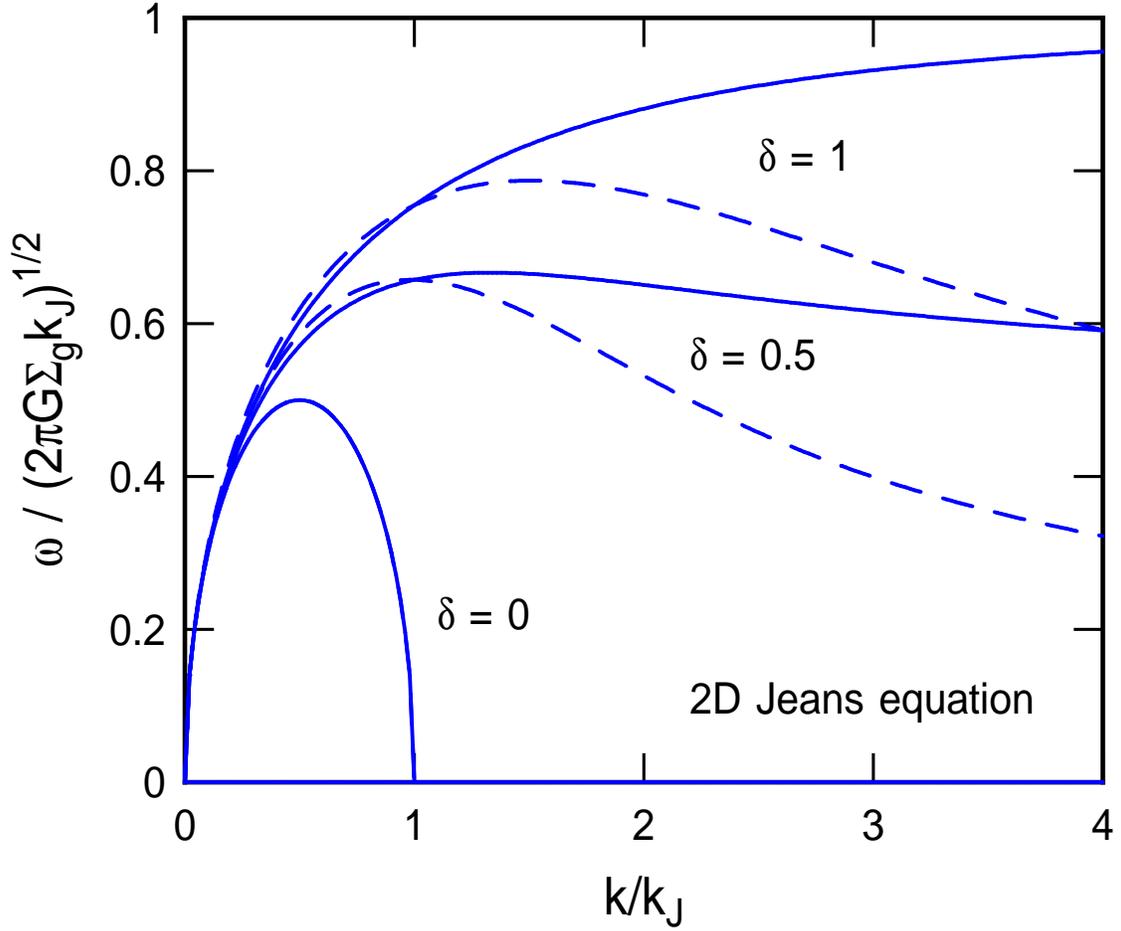} \caption{Dispersion relation for the
planar disk instability with only gas and no rotation, showing the
effect of turbulent dissipation. $\delta$ is the relative energy
dissipation rate compared to the perturbation crossing rate. Without
turbulent dissipation ($\delta=0$), the dispersion relation has the
usual upper limit for instability at the Jeans wavenumber $k=k_{\rm
J}$.  With turbulent dissipation, the instability also happens in
smaller regions because shocks remove energy and promote local
gravitational instabilities. Solid lines are for a velocity dispersion
that is independent of the size of the perturbation, as in the rest of
this paper, and dashed lines are for a velocity dispersion that
increases as the square root of the size of the perturbation.
}\label{jeans_w_delta}
\end{figure}

\clearpage
\begin{figure}
\epsscale{.9} \plotone{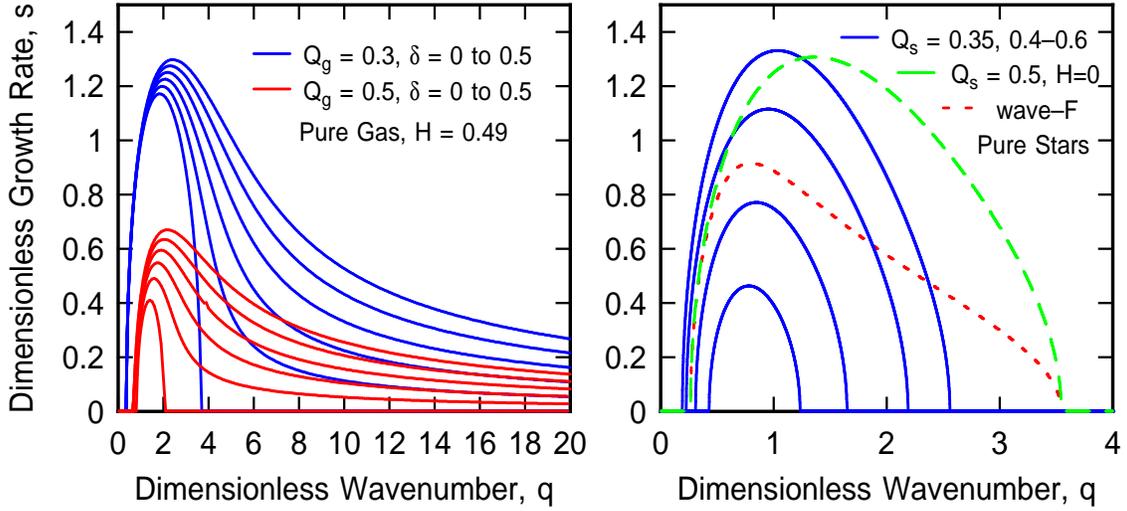} \caption{Solutions to the dispersion
relation between growth rate and wavenumber are shown for a pure gas
disk on the left and a pure star disk on the right. The different
curves of the same type and color in the left panel are for different
dimensionless dissipation rates. High dissipation leads to greater
instability at high wavenumber. On the right, the four blue curves
correspond to four values of $Q_{\rm s}$ including thickness
corrections. The dashed green curve is for a zero thickness disk. The
dotted red curve is a solution for a zero-thickness disk that uses a
reduction factor ${\cal F}$ evaluated in the case where the dispersion
relation is oscillatory; such oscillations are unphysical for the
assumed low $Q_{\rm s}$, because these would give growth solutions. In
comparison to the dashed green curve, the dotted red curve shows the
importance of using the appropriate expression for the reduction
factor. }\label{rafikov_dispersion_gasstars}
\end{figure}

\clearpage
\begin{figure}
\epsscale{.6} \plotone{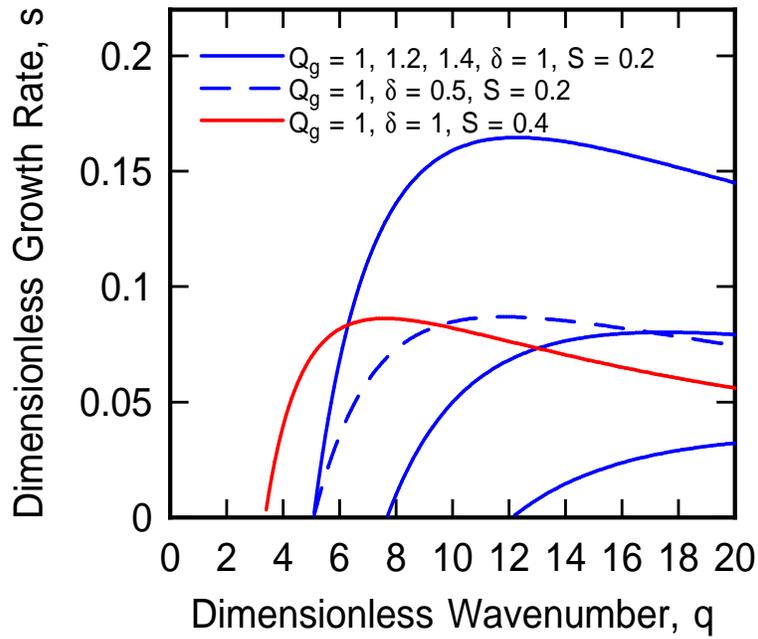} \caption{Dispersion relations for pure
gas disks are shown in cases formerly thought to be stable, $Q_{\rm
g}\ge1$. The destabilizing effect of the dissipation term involving
$\delta$ can be seen.  Thickness effects are included, and the
stability at low $q$ is from the dilution of in-plane self-gravity by
disk thickness. }\label{rafikov_dispersion_gas2}
\end{figure}

\clearpage
\begin{figure}
\epsscale{1.} \plotone{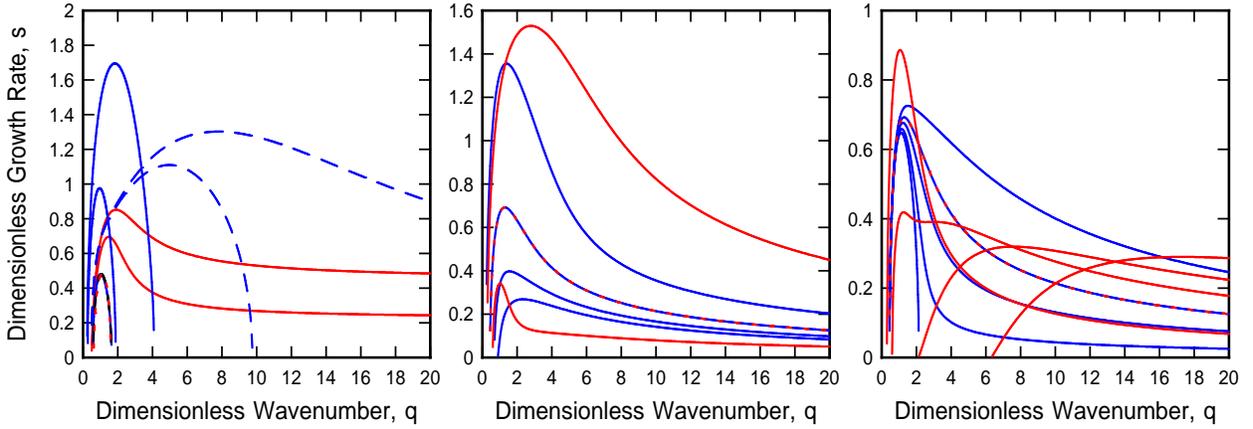} \caption{Solutions to the combined
gas+star dispersion relation are shown. The left hand panel has several
two-component cases that are effectively one-component, as a check on
the equations. The dashed blue curves are two-component with $Q_{\rm
s}=1$, $Q_{\rm g}=0.5$, and $S=0.5$, and with dissipation parameter
$\delta=0.5$ on the top and $\delta=0$ on the bottom. The middle panel
shows a range of values for $Q_{\rm s}$ and $Q_{\rm g}$ at fixed
$S=0.5$ and $\delta=0.5$. The sequence of blue curves with decreasing
height has $Q_{\rm s}=0.5$, 1, 1.5, and 2 for $Q_{\rm g}=1$, and the
sequence of red curves with decreasing height has $Q_{\rm g}=0.5$, 1,
and 1.5 for $Q_{\rm s}=1$. The right hand panel shows the effect of
dissipation as a sequence of blue curves ($\delta=0$, 0.1, 0.3, 0.5,
and 1 with $S=0.5$ from bottom to top) and the effect of the velocity
dispersion ratio as a sequence of red curves ($S=0.1$, 0.2, 0.3, 0.5,
and 1 with $\delta=0.5$ from right to left). On the right, all have
$Q_{\rm s}=Q_{\rm g}=1$. }\label{rafikov_dispersion_sg}
\end{figure}

\clearpage
\begin{figure}
\epsscale{1.} \plotone{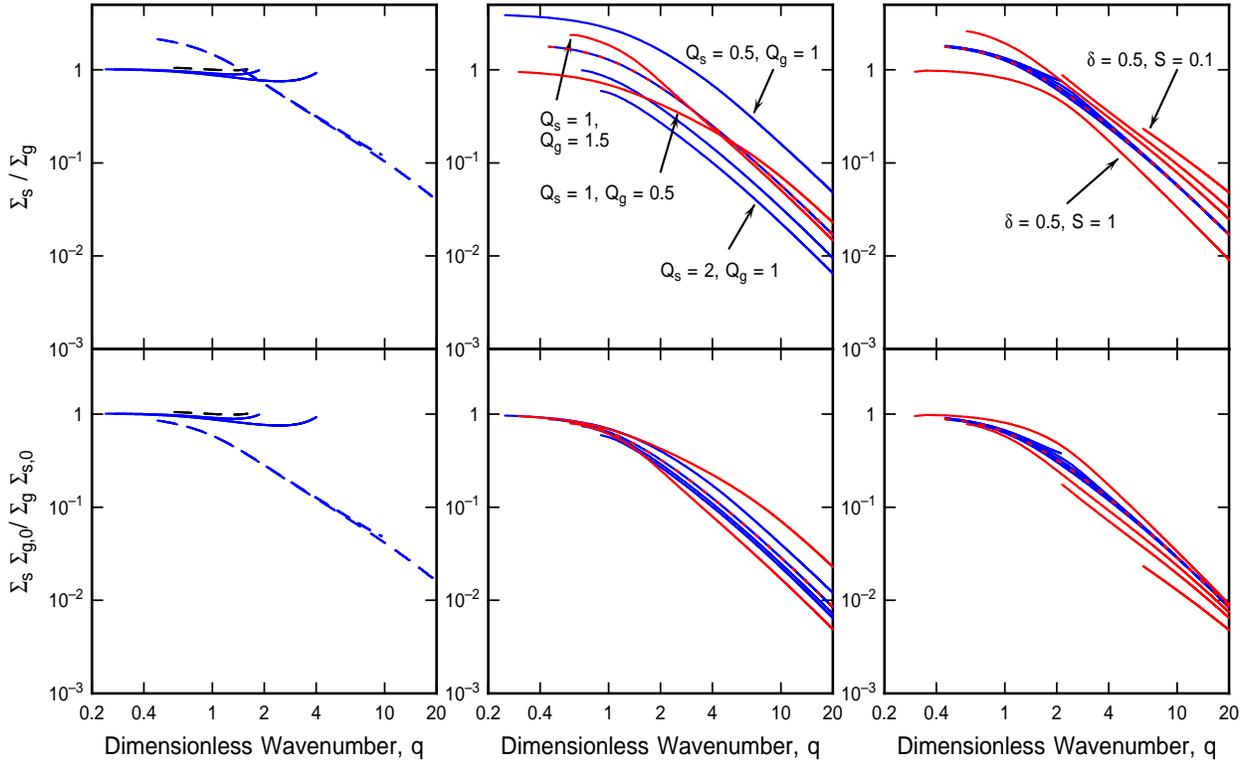} \caption{The ratio of perturbed stars to
gas (top) and this same ratio normalized to the initial star-to-gas
ratio (bottom) are shown versus wavenumber.  The order of the panels,
left to right, and the curve colors and types are the same as in Figure
7. }\label{rafikov_dispersion_sssg3}
\end{figure}

\clearpage
\begin{figure}
\epsscale{.7} \plotone{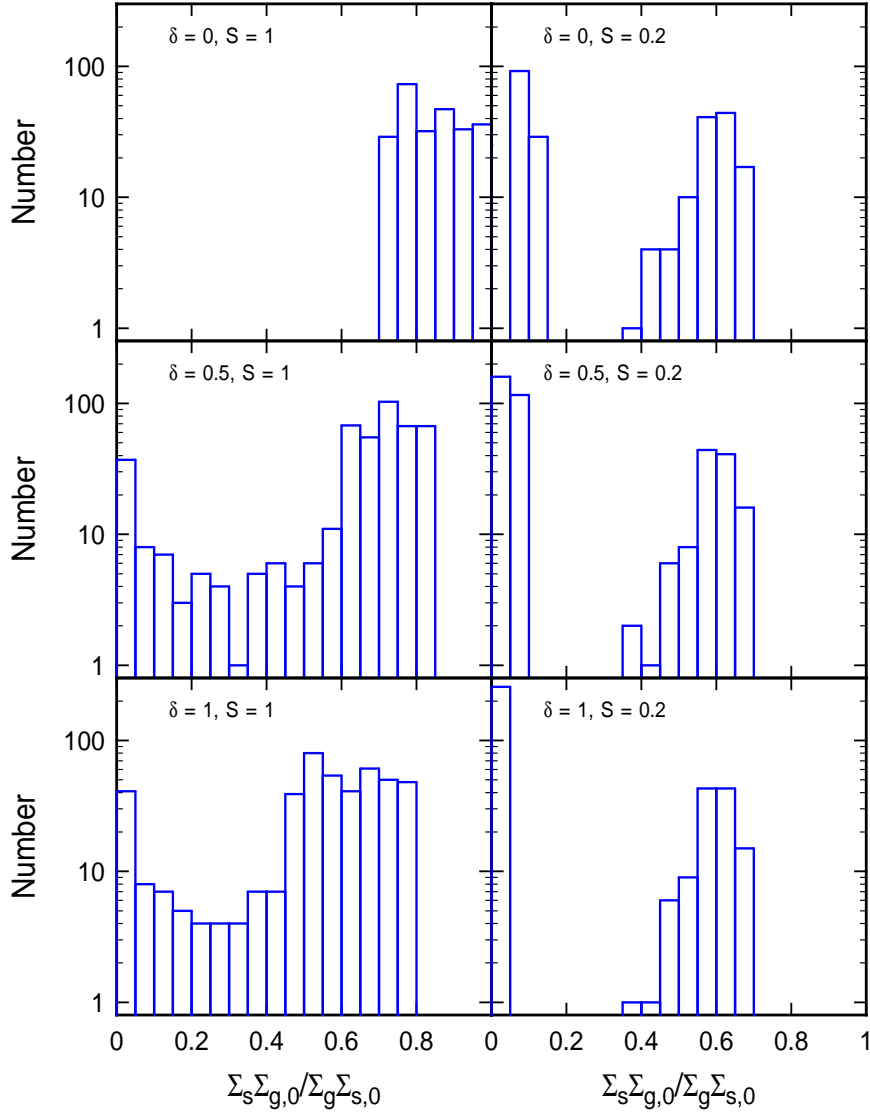} \caption{Histograms are shown of the
normalize ratio of perturbed stars to gas at the wavenumber of peak
growth. Each histogram is from a series of 800 cases with a range of
$Q_{\rm s}$ and $Q_{\rm g}$. Each panel has a different dissipation
parameter and velocity dispersion ratio, as indicated. As dissipation
increases in the high gas dispersion case (from top to bottom on the
left), the gas plays a more prominent role in the growth of the
instability (histograms shift to the left). }\label{rafikov_his}
\end{figure}

\clearpage
\begin{figure}
\epsscale{.7} \plotone{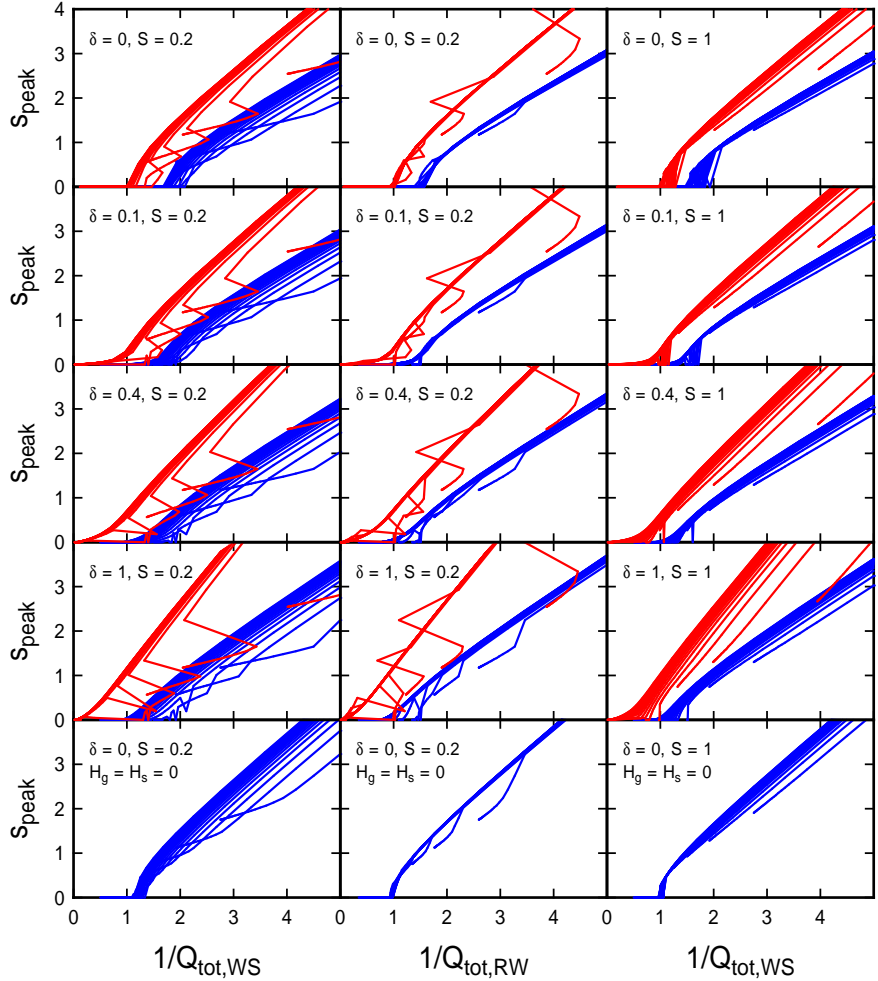} \caption{The peak growth rate of the
gas+star dispersion relation is plotted as a function of $1/Q_{\rm
tot}$ for the Wang \& Silk (1994) definition, on the left, and the
Romeo \& Wiegert (2011) definition in the middle. In these two cases,
the velocity dispersion ratio is 0.2. On the right, the velocity
dispersion ratio is unity and the two definitions of $Q_{\rm tot}$ are
the same because the disk is effectively single-component. Each panel
contains 20 blue curves and 20 red curves. Within each curve, $Q_{\rm
g}$ varies, while the different curves are for different $Q_{\rm s}$.
In the top four rows, blue curves use an expression for $Q_{\rm tot}$
in which thickness effects are not considered, even though they are
considered in the dispersion relation; red curves use expressions for
$Q_{\rm tot}$ that include thickness corrections, and are therefore
self-consistent.  The top four panels differ in their value of the
dissipation parameter $\delta$. The bottom row solves the dispersion
relation for a disk with zero thickness, which makes the red curves the
same as the blue curves. The top four rows show how increasing
dissipation always lowers $1/Q_{\rm tot}$ at the threshold for
instability, $s_{\rm peak}=0$. }\label{rafikov_s_vs_qgas}
\end{figure}

\clearpage
\begin{figure}
\epsscale{.6} \plotone{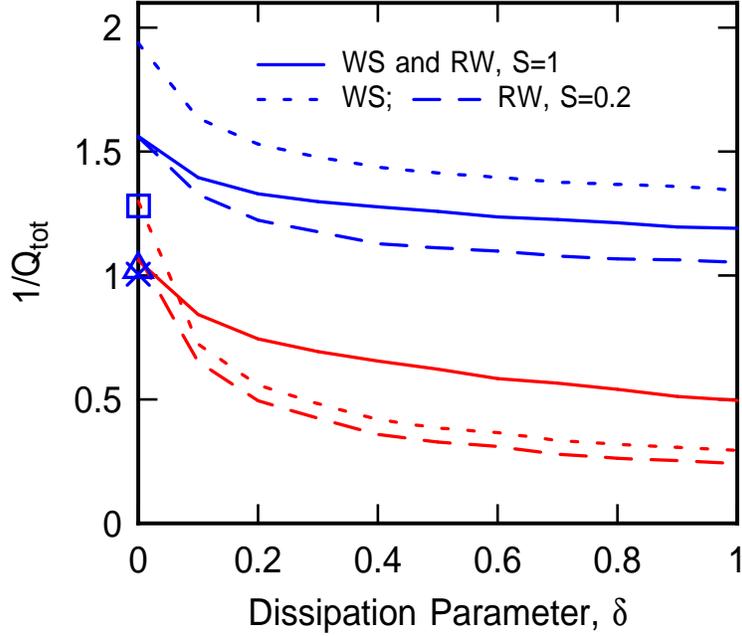} \caption{These curves plot the average
values of $1/Q_{\rm tot}$ where the peak growth rate, $s_{\rm peak}$,
is between 0.03 and 03, from Figure 10. These are essentially the
stability thresholds for the various definitions of $Q_{\rm tot}$. Blue
curves ignore thickness corrections in the definition of $Q_{\rm tot}$,
as in Figure 10, while red curves explicitly account for thickness in
$Q_{\rm tot}$ (all curves have thickness corrections in the actual
dispersion relation). The three symbols on the ordinate axis are cases
with zero-thickness disks.  Dissipation lowers $1/Q_{\rm tot}$ in all
cases. For reasonable values of the parameters, $Q_{\rm tot}$ has to
exceed 2 or 3 for a dissipative disk to become stable, instead of the
usual threshold of $\sim1$. }\label{rafikov_s_vs_qgas_zeros}
\end{figure}

\clearpage
\begin{figure}
\epsscale{.6} \plotone{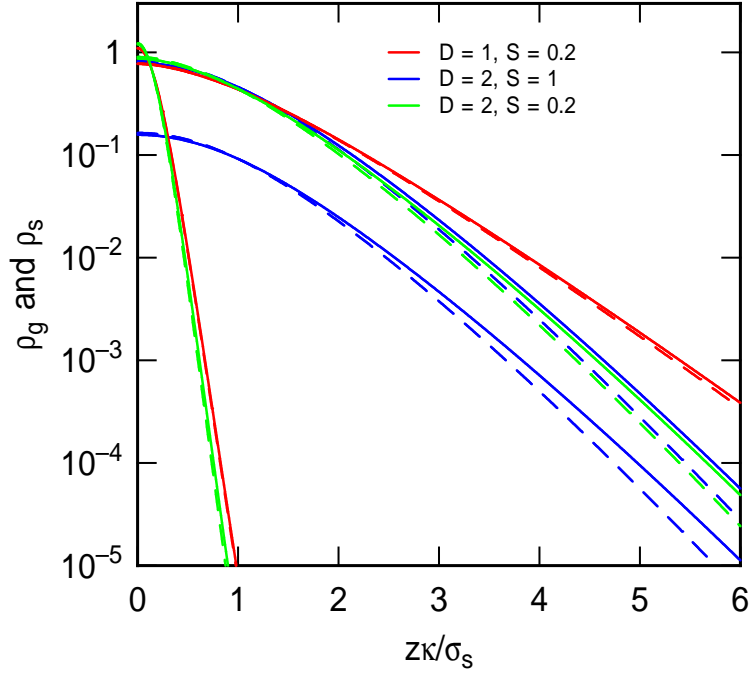} \caption{Vertical profiles for gas
(quickly falling curves) and stars for two values of the dimensionless
dark-matter parameter $D$ (defined above equation A10) and two values
of the ratio of gaseous to stellar velocity dispersion, $S$. The
abscissa is the dimensionless height above the midplane. Solid curves
are for a steep rotation curve, $\alpha=0.4$, and dashed curves are for
a nearly flat rotation curve, $\alpha=0.04$ (defined below equation
A8). Density is normalized to $\kappa^2/\left(2\pi GQ_{\rm s}\right)$
(see equations A11 and A12). These solutions assume $Q_{\rm s}=2$,
${\hat v}_{\rm rot}=8$, and $\Sigma_{\rm g,0}/\Sigma_{\rm s,0}=0.2$.
}\label{rafikov_jog}
\end{figure}

\clearpage
\begin{figure}
\epsscale{1.} \plotone{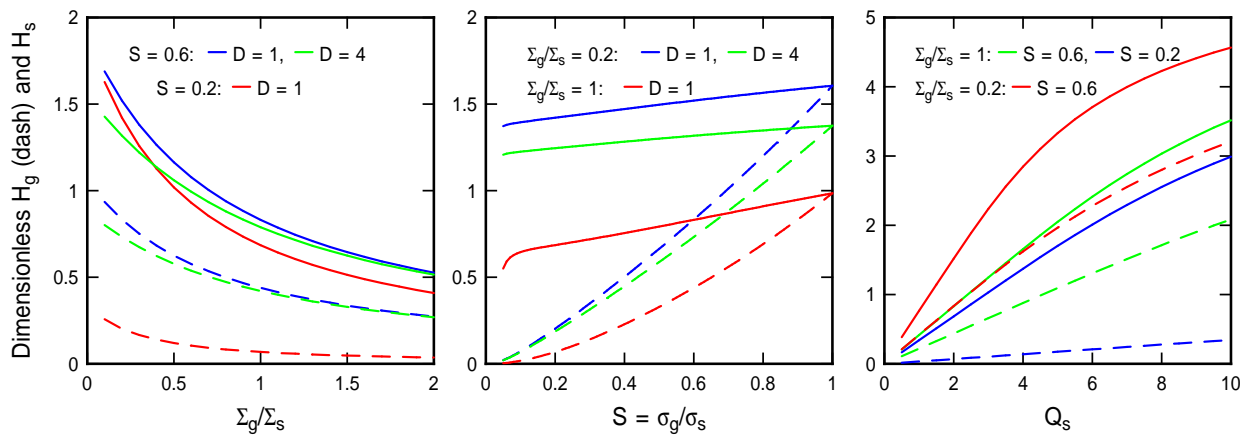} \caption{Dimensionless scale heights
are plotted versus the gas fraction in the disk on the left, versus the
gas-to-star velocity dispersion ratio in the middle, and versus $Q_{\rm
s}$ on the right. Parameters associated with the various colors are
indicated at the top of each panel. In the left and middle panels,
$Q_{\rm s}=2$. }\label{rafikov_jogh}
\end{figure}

\clearpage
\begin{figure}
\epsscale{.6} \plotone{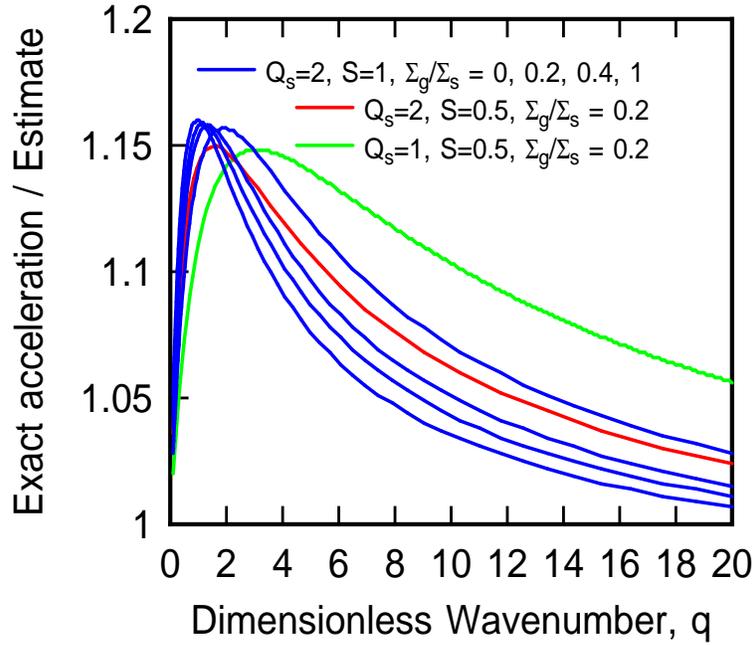} \caption{The curves show the correction
factors for the in-plane acceleration in a thick disk compared to the
approximate value using the term $(1+kH)$.  The correction factor for
the growth rate is approximately equal to the square root of these
acceleration factors. The correction factors are small, indicating that
the usual approximation for disk thickness is good.
}\label{rafikov_grav2}
\end{figure}

\end{document}